\documentclass{aa}

\usepackage{graphicx}
\usepackage{txfonts}
\usepackage{hyperref}
\begin{document} 

   \title{Trapping (sub-)Neptunes similar to TOI-216b at the inner disk rim}
   \subtitle{Implications for the disk viscosity and the Neptunian desert}
   
   \author{O. Chrenko\inst{1}
          \and
          R. O. Chametla\inst{1}
          \and
          D. Nesvorn\'{y}\inst{2}
          \and
          M. Flock\inst{3}
          }

   \institute{Charles University, Fac Math \& Phys, Astronomical Institute, V Hole\v{s}ovi\v{c}k\'{a}ch 747/2, 180 00 Prague 8, Czech Republic\\
   \email{chrenko@sirrah.troja.mff.cuni.cz}
   \and
   Department of Space Studies, Southwest Research Institute, 1050 Walnut St., Suite 300, Boulder, CO 80302, USA
   \and
   Max-Planck Institute for Astronomy, Königstuhl 17, D-69117 Heidelberg, Germany
   }
   
   \date{Received 9 July 2022 / Accepted 21 August 2022}

 
  \abstract
   {
   The occurrence rate of observed sub-Neptunes has a break at 0.1 au,
   which is often attributed to a migration trap at the inner rim
   of protoplanetary disks where a positive
   co-rotation torque prevents inward migration.
   }
   {
   We argue that conditions in inner disk regions are such that sub-Neptunes
   are likely to open gaps, lose the support of the co-rotation torque as their co-rotation regions become depleted, and
   the trapping efficiency then becomes uncertain.
   We study what it takes to trap such gap-opening planets at the inner disk rim.
   }
   {
   We performed 2D locally isothermal and non-isothermal hydrodynamic simulations of planet migration.
   A viscosity transition was introduced in the disk to
   (i) create a density drop and (ii) mimic the viscosity increase as the planet migrated
   from a dead zone towards a region with active magneto-rotational instability (MRI).
   We chose TOI-216b as a Neptune-like upper-limit test case, but we also explored different planetary
   masses, both on fixed and evolving orbits.
   }
   {
   For planet-to-star mass ratios $q\simeq(4$--$8)\times10^{-5}$, the density drop
   at the disk rim becomes reshaped due to a gap opening and is often
   replaced with a small density bump centred on the planet's co-rotation.
   Trapping is possible only if the bump retains enough gas mass and if the co-rotation region becomes azimuthally asymmetric, with an island of librating streamlines that accumulate a gas overdensity ahead of the planet.
   The overdensity exerts a positive torque that can counteract the negative torque of spiral arms. Under suitable conditions, the overdensity turns into a Rossby vortex. In our model, efficient trapping 
   depends on the $\alpha$ viscosity and its contrast across the viscosity transition.
   In order to trap TOI-216b, $\alpha_{\mathrm{DZ}}=10^{-3}$ in the dead zone requires $\alpha_{\mathrm{MRI}}\gtrsim5\times10^{-2}$
   in the MRI-active zone. If $\alpha_{\mathrm{DZ}}=5\times10^{-4}$, $\alpha_{\mathrm{MRI}}\gtrsim7.5\times10^{-2}$ is needed.
   }
   {
   We describe a new regime of a migration trap relevant for massive (sub-)Neptunes that puts valuable constraints on the levels of turbulent
   stress in the inner part of their natal disks.
   }

    
   \maketitle
%

\section{Introduction}
\label{sec:intro}

Statistical trends in the population of exoplanets tell us that
the most frequent planets on close-in orbits fall into the category
of super-Earths and sub-Neptunes \citep[with their radii between $1$ and $4\,R_{\oplus}$; e.g.][]{Mayor_etal_2011arXiv1109.2497M,Howard_etal_2012ApJS..201...15H,Petigura_etal_2013PNAS..11019273P,Winn_Fabrycky_2015ARA&A..53..409W}.
The occurrence rate of these planets increases sharply at
around $\sim$$0.1\,\mathrm{au}$ \citep[orbital periods $\sim$$10 \,\mathrm{d}$;][]{Petigura_etal_2018AJ....155...89P}, which is also
where the majority of inner planets in multi-planetary systems reside
\citep{Mulders_etal_2018AJ....156...24M}.

Due to the large amount of solids confined within super-Earths (compared to solar system
terrestrial planets) and due to the existence of H/He envelopes of sub-Neptunes \citep[e.g.][]{Hadden_Lithwick_2014ApJ...787...80H,Marcy_etal_2014PNAS..11112655M},
it is believed that their formation was likely finished before
their natal disk was dispersed \citep[e.g.][]{Lambrechts_etal_2019A&A...627A..83L,Ogihara_Hori_a_2020ApJ...892..124O,Ogihara_etal_2020ApJ...899...91O,Venturini_etal_2020A&A...643L...1V,Venturini_etal_2020A&A...644A.174V}. If so, these planets must have been subject to
disk-driven planetary migration which is often inward-directed \citep{Goldreich_Tremaine_1979ApJ...233..857G,Korycansky_Pollack_1993Icar..102..150K,Ward_1997Icar..126..261W,Miyoshi_etal_1999ApJ...516..451M,Masset_2002A&A...387..605M,Tanaka_etal_2002ApJ...565.1257T}
and thus an explanation is needed on how to stop the orbital migration and how to explain
the aforementioned clustering at 0.1 au. One possibility is the migration trap mechanism of
\cite{Masset_etal_2006ApJ...642..478M}. When a migrating planet encounters a surface density drop,
the component of the co-rotation torque driven by the vortensity gradient \citep[i.e. the horseshoe drag;][]{Ward_1991LPI....22.1463W} exhibits a positive boost that can counteract the negative
Lindblad torque of the spiral arms.
Consequently, the planetary orbit converges to a radius where the total torque is zero
and the migration effectively ceases. The clustering at 0.1 au might then be associated with the inner disk
rim\footnote{For the sake of clarity, let us point out that our definition of
the disk rim throughout the paper corresponds to the evaporation front of dust
grains that overlaps with the transition in the MRI activity of the disk. Our definition
should not be confused with the disk edge related to the magneto-spheric cavity.}
\citep[e.g.][]{Terquem_Papaloizou_2007ApJ...654.1110T,Izidoro_etal_2017MNRAS.470.1750I,Brasser_etal_2018ApJ...864L...8B} where a surface density drop is expected, perhaps due to a viscosity transition from the outer dead zone to the inner zone with active magneto-rotational instability \citep[MRI;][]{Flock_etal_2017ApJ...835..230F}.
\cite{Flock_etal_2019A&A...630A.147F} present a detailed physical model of the inner disk rim for a typical T Tauri star and predict a migration trap positioned near 0.1 au for a range of model parameters.

A specific example of an observed exoplanetary system that requires the trapping of the inner planet
is TOI-216 \citep{Dawson_etal_2019AJ....158...65D,Kipping_etal_2019MNRAS.486.4980K,Dawson_etal_2021AJ....161..161D}, which was studied in our recent work \cite{Nesvorny_etal_2022ApJ...925...38N}. The system is comprised of a sub-Solar K-type host star ($M_{\star}=0.77\,M_{\odot}$), an inner Neptune-class TOI-216b ($M_{\mathrm{p}}=0.059\,M_{\mathrm{Jup}}$, $P_{\mathrm{orb}}\simeq17.1\,\mathrm{d}$), and an outer half-Jupiter TOI-216c ($M_{\mathrm{p}}=0.56\,M_{\mathrm{Jup}}$, $P_{\mathrm{orb}}\simeq34.6\,\mathrm{d}$), with the planet pair captured firmly in the 2:1 mean-motion resonance. By examining various formation scenarios,
\cite{Nesvorny_etal_2022ApJ...925...38N} argue that trapping TOI-216b close to its observed orbital distance is an essential prerequisite for a successful creation of the 2:1 resonant lock and its subsequent (hydro)dynamical evolution towards the observed state (namely the eccentricity of TOI-216b and the libration amplitude in the resonance). Since the model of \cite{Flock_etal_2019A&A...630A.147F} can be applied to TOI-216 without significant changes,
\cite{Nesvorny_etal_2022ApJ...925...38N} conclude that the planet trap was likely facilitated by
the viscosity transition at the inner disk rim.
 
However, we also point out in \cite{Nesvorny_etal_2022ApJ...925...38N} that TOI-216b must have been capable of opening a relatively deep gap and thus it is unclear whether it could have been trapped at
the inner disk rim efficiently.
The reason for this doubt is that if a planet opens a gap, its co-rotation region becomes emptied 
and thus the contribution of the positive co-rotation torque is diminished.
A reliable answer can only be provided by means of detailed hydrodynamic simulations which were
not the main objective of \cite{Nesvorny_etal_2022ApJ...925...38N} and we thus focus on them here.

Although we chose TOI-216b as our reference case, the efficiency of the trapping mechanism
is a question that applies to sub-Neptunes in general. To demonstrate that,
let us recall that the ability of a planet
to open a gap and thus switch from the Type I to Type II migration regime predominantly depends on three quantities:
the planet-to-star mass ratio $q$, the disk alpha viscosity $\alpha$, and the disk aspect ratio $h$.
One of the available predictions for the gap depth \citep{Kanagawa_etal_2018ApJ...861..140K} leads to
\begin{equation}
    \frac{\Sigma_{\mathrm{gap}}}{\Sigma_{0}} = \frac{1}{1+0.04K} \, ,
    \label{eq:kanagawa}
\end{equation}
where $\Sigma_{\mathrm{gap}}$ is the minimal surface density at the gap bottom, $\Sigma_{0}$ is the unperturbed surface density, and 
\begin{equation}
    K = q^{2}h^{-5}\alpha^{-1} \, .
    \label{eq:K}
\end{equation}
The key fact to take into account is that $h$ is very small at 0.1 au. \cite{Flock_etal_2019A&A...630A.147F}
find $h\simeq0.023$ near the viscosity transition and this value is relatively well
constrained because it is simply a result of the thermal balance in a stellar-irradiated disk\footnote{To increase $h$, one would need to consider a more luminous irradiating star. However, that would shift the whole rim farther out, making it inconsistent with the expected location of the migration trap.}.
Before reaching the inner rim, the planet is embedded in the dead zone
where $\alpha=10^{-4}$--$10^{-3}$ can be expected, with
the lower limit corresponding, for example, to a hydrodynamic turbulence \citep{Nelson_etal_2013MNRAS.435.2610N,Klahr_Hubbard_2014ApJ...788...21K}
and the upper limit arising from magneto-hydrodynamic (MHD) models \citep{Flock_etal_2017ApJ...835..230F}.
Choosing the planet-to-star mass ratio $q=3\times10^{-5}$ (corresponding to a ten-Earth-mass planet
orbiting a Solar-mass star), one obtains\footnote{For a comparison, TOI-216b has $q=7.3\times10^{-5}$ which would lead to $\Sigma_{\mathrm{gap}}/\Sigma_{0}\simeq0.003$--$0.029$ with other parameters unchanged.} $\Sigma_{\mathrm{gap}}/\Sigma_{0}\simeq0.018$--$0.15$.
Such an excavation of the co-rotation region would affect the co-rotation torque without a doubt \citep{Kanagawa_etal_2018ApJ...861..140K}.
However, it is not a priori clear if Eqs.~(\ref{eq:kanagawa}) and (\ref{eq:K})
still apply once the planet reaches the viscosity transition
as they were derived for constant $\alpha$.
Our simulations are thus meant to clarify the picture.

To date, there have not been a large number of hydrodynamic studies dealing with the migration
of planets that open moderately deep gaps across the viscosity transition of the inner rim. In \cite{Ataiee_Kley_2021A&A...648A..69A}, the planet was either too massive (i.e. having a Jupiter mass)
or the aspect ratio was too large (typically $h=0.05$; the value of $h=0.03$ was only considered in a limited number of cases). \cite{Romanova_etal_2019MNRAS.485.2666R} performed detailed 3D simulations at the disk-cavity boundary with the main parameters $h=0.03$, $q=1.5\times10^{-5}$ or $4.5\times10^{-5}$,
and zero viscosity. They find very efficient trapping, but their simulation timescales
were relatively short and the planet was allowed to migrate even before the (possible) gap
was relaxed (see their Fig. B2). Moreover, they did not consider a viscosity transition.
Finally, \cite{Faure_Nelson_2016A&A...586A.105F} studied the planet migration at the inner
edge of a dead zone in the presence of a cyclically evolving vortex. For their setup, they
identify $q\in(3\times10^{-5},10^{-4})$ as the optimal range for efficient trapping.
Based on the aforementioned works, it is not straightforward to assess whether
(sub-)Neptunes are subject to efficient trapping at the disk edge,
and re-investigating the migration trap in this planet mass range is thus worthwhile.

At the same time, the trap should not be
too efficient when $q$ increases towards super-Neptunes because
the exoplanet occurrence rate steeply decreases already around the Neptune mass and results in the Neptunian desert.
Having an efficient migration trap in this mass range might result in an inconsistency
between the planet migration theory and observations.
We think that this is an important dynamical aspect and it is further elaborated
on throughout the paper. We point out, however, that our study cannot rule out other mechanisms
that might have created the Neptunian desert, such as photoevaporation \citep[e.g.][]{Owen_Wu_2013ApJ...775..105O,Lopez_Fortney_2013ApJ...776....2L,Owen_Lai_2018MNRAS.479.5012O,McDonald_etal_2019ApJ...876...22M,Hallatt_Lee_2022ApJ...924....9H}, tidal stripping \citep[e.g.][]{Matsakos_Konigl_2016ApJ...820L...8M,Owen_Lai_2018MNRAS.479.5012O},
or pebble isolation
\citep[e.g.][]{Bitsch_2019A&A...630A..51B,Venturini_etal_2020A&A...644A.174V}.

The paper is structured as follows. Our numerical models are described in Sect.~\ref{sec:methods}.
Sect.~\ref{sec:simu_iso} is devoted to 2D locally isothermal simulations with the viscosity
transition at 1 au which enable us to explore a larger portion of the parametric space 
and get a basic grasp of the processes that play a decisive role for the efficiency of the
migration trap. In Sect.~\ref{sec:simu_noniso}, we perform more realistic non-isothermal
simulations at 0.1 au to finalize our findings. A general discussion is provided
in Sect.~\ref{sec:discussion} and Sect.~\ref{sec:conclusions} summarizes our conclusions.

\section{Methods}
\label{sec:methods}
   
We used the Fargo3D code \citep{Benitez-Llambay_Masset_2016ApJS..223...11B},
which is a finite-difference solver based on upwind methods \citep{Stone_Norman_1992ApJS...80..753S},
paired with the fast orbital advection algorithm \citep{Masset_2000A&AS..141..165M}.
We accounted for a viscous protoplanetary disk (its gaseous component),
a central star, and a single planet embedded in the disk.
The system was evolved on a 2D polar staggered mesh
characterized by the radius $r$ and azimuth $\theta$.
The frame was centred on the host star, but co-rotated
with the planet.

\subsection{Locally isothermal 2D model}
\label{sec:model_iso}

Governing equations of the model read
as follows:\begin{equation}
    \frac{\partial\Sigma}{\partial t} + \nabla\cdot\left(\Sigma\vec{v}\right) = 0 \, ,
    \label{eq:conti}
\end{equation}
\begin{equation}
    \frac{\partial\vec{v}}{\partial t} + \left(\vec{v}\cdot\nabla\right)\vec{v} = - \frac{1}{\Sigma}\nabla P + \frac{1}{\Sigma}\nabla\cdot\tens{T} - \nabla\phi \, ,
    \label{eq:ns}
\end{equation}
where $\Sigma$ is the gas surface density, $\vec{v}$ is the velocity vector,
$P$ is the pressure, $\tens{T}$ is the viscous stress tensor, and $\phi$ is the gravitational
potential. The equation of state is given by
\begin{equation}
    P = c_{\mathrm{s}}^{2}\Sigma = H^{2}\Omega_{\mathrm{K}}^{2}\Sigma \, ,
    \label{eq:eos_iso}
\end{equation}
where $c_{\mathrm{s}}$ is the sound speed, $\Omega_{\mathrm{K}}$ is the Keplerian frequency,
and $H$ is the pressure scale height. The latter is a radially fixed function of the aspect ratio $h=H/r$,
which was uniform in our locally isothermal model (i.e. the disk flaring was neglected).

The potential component related to the planet was smoothed via
\begin{equation}
    \phi_{\mathrm{p}} = -\frac{GM_{\mathrm{p}}}{\sqrt{d^{2}+r_{\mathrm{sm}}^{2}}} \, ,
    \label{eq:pot_pl}
\end{equation}
where $d$ is the distance from a grid cell to the planet
and $r_{\mathrm{sm}}=0.6H$ is the smoothing length that allows one to avoid numerical divergence at $d=0$ and also leads to disk torques that resemble those found in 3D simulations \citep{Muller_etal_2012A&A...541A.123M}. The gravitational potential also included disk- and planet-driven indirect terms 
that arise from the usage of a non-inertial reference frame.

Since we are mostly concerned with planets that open (at least partial) gaps, it is likely
that these planets also accumulate gas in a circumplanetary disk. However, this region is under-resolved
and due to the lack of self-gravity, its contribution to the torque exerted on the planet
is incorrect. Therefore, it is customary to exclude a part of the gas contained within the Hill sphere
when calculating the disk torque
(but not when computing the back-reaction of the planet on the gas).
When evaluating the gravitational acceleration exerted by a gas cell on the planet,
we multiplied by the tapering formula adopted from \cite{Crida_etal_2008A&A...483..325C}:
\begin{equation}
    f_{\mathrm{cut}} = \left[\exp{\left(-\frac{d/R_{\mathrm{H}}-p}{p/10}+1\right)}\right]^{-1} \, ,
    \label{eq:hillcut}
\end{equation}
where $p=0.8$. In addition, we enabled the module of Fargo3D that activates the correction for the shift 
of Lindblad resonances that is also related to the lack of self-gravity \citep{Baruteau_Masset_2008ApJ...678..483B}.
The location of resonances is improved by subtracting the azimuthally averaged gas density
from $\Sigma$ prior to the torque evaluation.

The disk was initialized with a uniform mass accretion rate
\begin{equation}
    \dot{M} = 3\pi\nu\Sigma \, ,
    \label{eq:mdot}
\end{equation}
where the kinematic viscosity $\nu$ was calculated using the alpha prescription \citep{Shakura_Sunyaev_1973A&A....24..337S}
\begin{equation}
    \nu = \alpha \frac{c_{\mathrm{s}}^{2}}{\Omega_{\mathrm{K}}} \, .
    \label{eq:nu}
\end{equation}
To mimic the viscosity transition and the surface density drop at the inner
disk rim, we used \citep[for a similar temperature-dependent transition, see Eq.~\ref{eq:alpha_Tdep} or][]{Flock_etal_2016ApJ...827..144F}
\begin{equation}
    \alpha(r) = \frac{1}{2}\left(\alpha_{\mathrm{MRI}}-\alpha_{\mathrm{DZ}}\right)\left(1-\tanh{\frac{r-r_{\mathrm{MRI}}}{\Delta r_{\mathrm{MRI}}}}\right) + \alpha_{\mathrm{DZ}} \, ,
    \label{eq:transition}
\end{equation}
where $\alpha_{\mathrm{MRI}}$ accounts for the alpha viscosity in the MRI-active zone
inwards from the transition radius $r_{\mathrm{MRI}}$ and $\alpha_{\mathrm{DZ}}$ characterizes
the viscous stress in the dead zone at $r>r_{\mathrm{MRI}}$. The sharpness of the viscosity 
transition was controlled by the parameter $\Delta r_{\mathrm{MRI}}$.

The radial velocity field was initially
\begin{equation}
    v_{r} = - \frac{3}{2}\frac{\nu}{r} \, ,
    \label{eq:v_r}
\end{equation}
and the azimuthal velocities $v_{\theta}$ were set by the centrifugal
balance between the gravity of the star and the pressure support within the disk.
The initial conditions for $\Sigma$, $v_{r}$ and $v_{\theta}$ were also used as boundary conditions.
To provide a smooth transition between the boundaries and the perturbed
disk, we damped $\Sigma$ and $v_{r}$ towards their initial values within
radially narrow zones adjacent to the boundaries. The damping recipe was that of \cite{deValBorro_etal_2006MNRAS.370..529D}.

\subsection{Non-isothermal 2D model}
\label{sec:model_noniso}

\begin{table}[!t]
    \caption{Parameters varied in our simulations.}
    \centering
    \begin{tabular}{ll}
    \hline \hline
        Parameter & Notation \\
    \hline
        Planet-to-star mass ratio & $q$ \\
        MRI-active viscosity & $\alpha_{\mathrm{MRI}}$ \\
        Dead-zone viscosity & $\alpha_{\mathrm{DZ}}$ \\
        Mass accretion rate & $\dot{M}$ \\
        Transition width parameter & $\Delta r_{\mathrm{MRI}}$ \\
        \hline
    \end{tabular}
    \label{tab:params_varied}
\end{table}
   
\begin{table}[!t]
    \caption{Fixed parameters of the locally isothermal simulations.}
    \centering
    \begin{tabular}{lll}
        \hline \hline
        Parameter & Notation & Value \\
    \hline
         Grid size & $N_{r}\times N_{\theta}$ & $500 \times 1536$ \\
         Inner radial boundary & $r_{\mathrm{in}}$ & $0.5\,\mathrm{au}$ \\
         Outer radial boundary & $r_{\mathrm{out}}$ & $3.6\,\mathrm{au}$ \\
         Aspect ratio & $h$ & $0.023$ \\
         Viscosity transition radius & $r_{\mathrm{MRI}}$ & $1\,\mathrm{au}$ \\
         Stellar mass & $M_{\star}$ & $0.77\,M_{\odot}$ \\
         Time measure & $P_{\mathrm{orb}}(r_{\mathrm{MRI}})$ & $\simeq$$416\,\mathrm{d}$ \\ 
         \hline
    \end{tabular}
    \label{tab:params_iso}
\end{table}

\begin{table}[!t]
    \caption{Fixed parameters of the non-isothermal simulations.}
    \centering
    \begin{tabular}{lll}
        \hline \hline
        Parameter & Notation & Value \\
    \hline
         Grid size & $N_{r}\times N_{\theta}$ & $320 \times 1156$ \\
         Inner radial boundary & $r_{\mathrm{in}}$ & $0.07\,\mathrm{au}$ \\
         Outer radial boundary & $r_{\mathrm{out}}$ & $0.4\,\mathrm{au}$ \\
         Viscosity transition radius & $r_{\mathrm{MRI}}$ & $0.11\,\mathrm{au}$ \\
         Stellar mass & $M_{\star}$ & $0.77\,M_{\odot}$ \\
         Time measure & $P_{\mathrm{orb}}(r_{\mathrm{MRI}})$ & $\simeq$$15\,\mathrm{d}$ \\
         Stellar temperature & $T_{\star}$ & $4300\,\mathrm{K}$ \\
         Stellar radius & $R_{\star}$ & $2.4\,R_{\odot}$ \\
         Opacity & $\kappa$ & $1.3\,\mathrm{cm}^{2}\,\mathrm{g}^{-1}$ \\
         Photospheric height at $r>r_{\mathrm{cond}}$ & $f_{\mathrm{ph}}$ & $3.3H$ \\
         Photospheric height at $r_{\mathrm{cond}}$ & $f_{\mathrm{ph,cond}}$ & $4.5H$ \\
         Surface temperature at $r_{\mathrm{cond}}$ & $T_{\mathrm{cond}}$ & $1150\,\mathrm{K}$ \\
         Evaporation temperature & $T_{\mathrm{ev}}$ & $1370\,\mathrm{K}$ \\
         Emission-to-absorption efficiency & $\epsilon$ & 7/13 \\
         Mean molecular weight & $\mu$ & $2.3$ \\
         Adiabatic index & $\gamma$ & $1.43$ \\
            \hline
    \end{tabular}
    \label{tab:params_noniso}
\end{table}

We relaxed the locally isothermal approximation from previous Sect.~\ref{sec:model_iso}
by replacing Eq.~(\ref{eq:eos_iso}) with
\begin{equation}
    P = (\gamma-1)\mathcal{E} \, ,
    \label{eq:eos_noniso}
\end{equation}
where $\gamma$ is the adiabatic index and $\mathcal{E}$ is the internal energy density. To describe
the temporal evolution of $\mathcal{E}$, we followed the one-temperature approximation 
(i.e. a thermodynamic equilibrium between the gas and photons of thermal radiation)
and considered \citep[e.g.][]{Kley_Crida_2008,Pierens_2015MNRAS.454.2003P,Chrenko_etal_2017A&A...606A.114C,Ziampras_etal_2020A&A...637A..50Z}
\begin{equation}
    \frac{\partial\mathcal{E}}{\partial t} + \nabla\cdot\left(\mathcal{E}\vec{v}\right) = - P\nabla\cdot\vec{v} + Q_{\mathrm{visc}} + Q_{\mathrm{irr}} - Q_{\mathrm{vert}} \, ,
    \label{eq:energy}
\end{equation}
where $Q_{\mathrm{visc}}$ is the viscous heating term, $Q_{\mathrm{irr}}$ is the heating from stellar
irradiation, $Q_{\mathrm{vert}}$ is the vertical cooling by radiation escape from the disk,
and the term $-P\nabla\cdot\vec{v}$ accounts for the compressional heating. Let us note
that although our implementation allows for in-plane radiative diffusion, it was not
considered in this work and therefore omitted in Eq.~(\ref{eq:energy}).

The closure relation for viscous heating is given in Appendix~\ref{sec:app_visc} (Eq.~\ref{eq:qvisc}).
The irradiation-cooling balance can be written as
\begin{equation}
    Q_{\mathrm{irr}} - Q_{\mathrm{vert}} = \frac{2\sigma}{\tau_{\mathrm{eff}}}\left(T_{\mathrm{irr}}^{4}-T^{4}\right) \, ,
    \label{eq:irr_vs_cool}
\end{equation}
where $\sigma$ is the Stefan-Boltzmann constant,
$T_{\mathrm{irr}}$ is the midplane temperature of a passive disk,
and $T$ is the instantaneous temperature related to the internal energy density
as $\mathcal{E}=\rho c_{V} T$, $c_{V}$ being the heat capacity at constant volume.
To calculate the effective optical depth $\tau_{\mathrm{eff}}$ which allowed us to estimate
local radiative losses, we followed \cite{Hubeny_1990ApJ...351..632H} and \cite{DAngelo_Marzari_2012ApJ...757...50D}:
\begin{equation}
    \tau_{\mathrm{eff}} = \frac{3}{8}\tau_{\mathrm{opt}} + \frac{1}{2} + \frac{1}{4\tau_{\mathrm{opt}}} \, ,
    \label{eq:tau_eff}
\end{equation}
with
\begin{equation}
    \tau_{\mathrm{opt}} = \frac{1}{2}\kappa\Sigma \, .
    \label{eq:tau_opt}
\end{equation}
In writing Eq.~(\ref{eq:tau_opt}), we introduced a single grey opacity $\kappa$,
which was considered uniform.

Ultimately, we aimed for our model to be applicable at the inner rim
of protoplanetary disks, which is thermodynamically complex. \cite{Flock_etal_2016ApJ...827..144F,Flock_etal_2017ApJ...835..230F} performed vertically
resolved radiation (magneto)hydrodynamic studies of the inner rim structure
and identified four distinct regions when moving from the star towards larger radii
(see also Fig.~\ref{fig:noniso_disk}):
(A) an optically thin region of pure gas, (B) an optically thin halo with
a low dust concentration, (C) a rounded off condensation front of dust grains that
is optically thick to stellar irradiation \citep[and can be followed by a radially narrow self-shadowed region;][]{Flock_etal_2019A&A...630A.147F}, and
(D) a standard flaring disk with the photospheric height of several $H$.
The viscosity transition at the inner rim is related to the ionization temperature that triggers the MRI
and it is often located in the proximity of the boundary between regions C and D.
Therefore, our goal was to reproduce the thermal structure of regions C and D\footnote{Extending our model towards regions B and A would be challenging because the disk needs to be optically thick to its own thermal emission in the unresolved vertical direction for the one-temperature approximation and also for Eqs.~(\ref{eq:irr_vs_cool}) and (\ref{eq:tau_eff}) to remain valid.}
because they contain the first possible trap\footnote{Obviously, there could be other traps farther out in the disk but we assume that the planet managed to reach the rim or was formed in situ.}
that an inward-migrating planet encounters when spiraling towards the star.

Although our model is not vertically resolved, it is possible to follow the approach of \cite{Ueda_etal_2017ApJ...843...49U}, who analytically studied the thermal balance found in numerical simulations of \cite{Flock_etal_2016ApJ...827..144F},
and recover the temperature profile in regions C and D with a satisfactory precision (as we show in Sect.~\ref{sec:simu_noniso} and Fig.~\ref{fig:noniso_disk}). We took
\begin{equation}
T_{\mathrm{irr}} = \max\left(T_{\mathrm{trans}},T_{\mathrm{thick}}\right) \, ,
\label{eq:Tirr}
\end{equation}
where $T_{\mathrm{trans}}$ is the temperature in region C and $T_{\mathrm{thick}}$ is the temperature
in region D. For the former, \cite{Ueda_etal_2017ApJ...843...49U} find
\begin{equation}
T_{\mathrm{trans}}^{4} = \frac{1}{2\pi}\left[\arctan\left(\frac{r_{\mathrm{cond}}-r}{0.05r_{\mathrm{cond}}}\right)+\frac{\pi}{2}\right]T_{\mathrm{cond}}
,\end{equation}
where $r_{\mathrm{cond}}$ is the radius of dust condensation at all heights above the midplane and
$T_{\mathrm{cond}}$ is the temperature of the front of fully condensed dust.
The remaining closure relations are
\begin{equation}
    r_{\mathrm{cond}} = r_\mathrm{rim}\sqrt{1+\Gamma} \, ,
    \label{eq:r_cond}
\end{equation}
where 
\begin{equation}
    r_{\mathrm{rim}} = \frac{1}{2}\sqrt{\frac{1}{\epsilon}}\left(\frac{T_{\star}}{T_{\mathrm{ev}}+100\,\mathrm{K}}\right)^{2}R_{\star} \, ,
    \label{eq:r_rim}
\end{equation}
and
\begin{multline}
    \Gamma = 3.1\left(\frac{r_{\mathrm{rim}}}{0.46\,\mathrm{au}}\right)^{-12/7}\left(\frac{f_{\mathrm{ph,cond}}}{4.8}\right)^{8/7}\left(\frac{T_{\star}}{10^{4}\,\mathrm{K}}\right)^{32/7} \\
    \times\left(\frac{M_{\star}}{2.5M_{\odot}}\right)^{-1/2}\left(\frac{R_{\star}}{2.5R_{\odot}}\right)^{16/7} \, .
    \label{eq:gamma_trans}
\end{multline}
In these expressions, $r_{\mathrm{rim}}$ is the innermost radius where the disk midplane becomes thick
to stellar irradiation, $\epsilon$ is the emission-to-absorption efficiency of dust grains, $T_{\star}$ is the stellar temperature, $T_{\mathrm{ev}}$ is the temperature in the dust 
halo where the grains are evaporating (region B), $R_{\star}$ is the stellar radius, $f_{\mathrm{ph,cond}}$
is the photospheric height at $r_{\mathrm{cond}}$ (given in multiples of $H$), and $M_{\star}$ is the stellar mass.

For the temperature in region D, we derive a slightly modified formula with respect to \cite{Ueda_etal_2017ApJ...843...49U} in Appendix~\ref{sec:app_irr}:
\begin{equation}
    T_{\mathrm{thick}}^{4} = \frac{1}{2}\left[\frac{4}{3\pi\epsilon}\left(\frac{R_{\star}}{r}\right)^{3}T_{\star}^{4} + \left(0.262\frac{f_{\mathrm{ph}}}{\epsilon}R_{\star}^{2}\sqrt{\frac{\left(\gamma-1\right)c_{V}}{GM_{\star}r^{3}}}T_{\star}^{4}\right)^{8/7}\right] \, ,
    \label{eq:Tthick}
\end{equation}
where $f_{\mathrm{ph}}$ is the photospheric height in region D (again given in multiples of H).

In order to mimic the decrease of the vertical optical depth when moving from region D to region C,
we introduced an ad hoc modification of $\tau_{\mathrm{opt}}$ by multiplying it with
\begin{equation}
    f_{\tau} = \frac{1}{2}\left(f_{\tau\mathrm{,trans}}-f_{\tau\mathrm{,thick}}\right)\left(1-\tanh{\frac{r-r_{\mathrm{cond}}}{0.05r_{\mathrm{cond}}}}\right) + f_{\tau\mathrm{,thick}} \, ,
    \label{eq:tau_drop}
\end{equation}
where $f_{\tau\mathrm{,trans}}=0.1$ and $f_{\tau\mathrm{,thick}}=1$.
Equipped with all the formulae for the heating and cooling terms,
Eq.~(\ref{eq:energy}) was solved for temperature $T$ in an implicit form
using $\mathcal{E}=\rho c_{V} T$. As we omitted the in-plane radiative diffusion,
the implicit update has a simple analytic solution that can be performed
cell by cell.

In the non-isothermal model, we used the adiabatic versions
of the sound speed and pressure scale height:
\begin{equation}
    H = \frac{c_{\mathrm{s}}}{\sqrt{\gamma}\Omega_{\mathrm{K}}} = \frac{\sqrt{\gamma P/\Sigma}}{\sqrt{\gamma}\Omega_{\mathrm{K}}} \, .
    \label{eq:H_noniso}
\end{equation}
All planet-related calculations and the smoothed planetary potential remained the same
as in Sect.~\ref{sec:model_iso},
the only difference was that we used the azimuthally averaged $H$ when computing the potential smoothing length.

Eqs.~(\ref{eq:mdot}--\ref{eq:v_r}) remain valid as well. But finding the initial state of the disk required
a more detailed approach because $\nu$ depends on the temperature via $c_{\mathrm{s}}$ and this
dependence propagates into the calculation of $\Sigma$ from $\dot{M}$. We thus proceeded iteratively:
Starting from an initial guess of the radial temperature profile, we calculated the density profile
from Eqs.~(\ref{eq:mdot}--\ref{eq:transition}). Then we kept the density fixed and we evolved
Eq.~(\ref{eq:energy}) over a time interval $\Delta t$ that was similar to the characteristic radiation
diffusion timescale in the disk. The steps were repeated until the relative change of $T$ and $\Sigma$ became smaller than $10^{-5}$. Subsequently, we confirmed that a correct equilibrium state was found by
propagating the full set of Eqs.~(\ref{eq:conti}), (\ref{eq:ns}) and (\ref{eq:energy}) over 1000 orbital periods $P_{\mathrm{orb}}$ (measured at the viscous transition). During the relaxation, the disk
was effectively kept in 1D by assuming an axial symmetry and setting the number of azimuthal sectors equal to one. After the relaxation, we prepared the disk for simulations with an embedded planet by 
expanding all hydrodynamic fields in azimuth and by converting velocities to the frame co-rotating with 
the planet.

The boundary conditions were constructed as follows. First, we used a power-law extrapolation of $T$ from
the active grid cells to the ghost cells\footnote{The temperature is not an independent variable of our model but we found the most stable behaviour of our boundary conditions when starting from $T$.}.
Next, we used $T$ in the ghost cells to find $\nu$. Afterwards, $v_{r}$
was set in the ghost cells using Eq.~(\ref{eq:v_r}). The calculation of $\Sigma$ differed for the inner and outer boundary in radius \citep[see also][]{Bitsch_etal_2014A&A...564A.135B}. At the inner boundary, we kept the inward-directed mass flux through the boundary uniform by using $\Sigma_{\mathrm{g}}=\Sigma_{\mathrm{a}}\nu_{\mathrm{a}}/\nu_{\mathrm{g}}$, where the subscripts `g' and `a' stand for the ghost and active cells, respectively. At the outer boundary, 
we used $\Sigma_{\mathrm{g}}=\dot{M}/(3\pi\nu_{\mathrm{g}})$ in order to replenish the disk material in a uniform way. Finally, $\mathcal{E}$ in the ghost cells was recovered from $T$ and $\Sigma$.

Damping zones were used in the non-isothermal model as well. During the relaxation stage, we damped only $v_{r}$
so that $\Sigma$ could adjust and reach an equilibrium. The target value of $v_{r}$, towards which
we damped, was again given by Eq.~(\ref{eq:v_r}), which depends on the instantaneous local temperature.
During the simulations
with an embedded planet, we damped $\Sigma$ and $v_{r}$ towards the equilibrium disk state found at the
end of the relaxation stage.

\section{Locally isothermal simulations at 1 au}
\label{sec:simu_iso}

\begin{figure}[!t]
    \centering
    \includegraphics[width=\columnwidth]{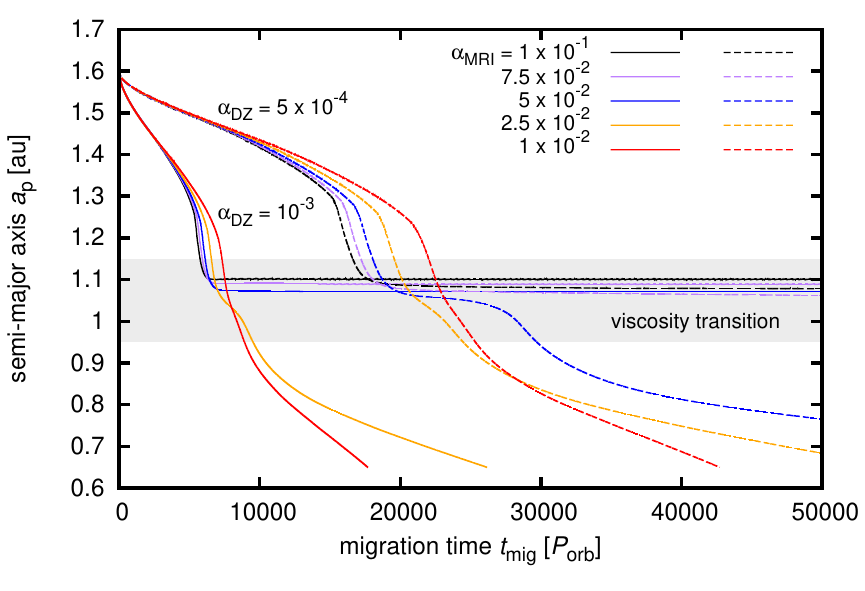}
    \caption{Temporal evolution of the semi-major axis $a_{\mathrm{p}}$ of migrating TOI-216b.
    Horizontal axis measures the migration time $t_{\mathrm{mig}}$ (we do not show the
    simulation stages during which the planet was non-migrating). Solid lines are for 
    $\alpha_{\mathrm{DZ}}=10^{-3}$ and dashed lines are for $\alpha_{\mathrm{DZ}}=5\times10^{-4}$.
    Different values of $\alpha_{\mathrm{MRI}}$ are distinguished by colour and specified in the
    plot legend. The shaded area marks the disk region over which $\alpha_{\mathrm{DZ}}$ increases
    to $\alpha_{\mathrm{MRI}}$. Outwards from the shaded area, there is the dead zone.
    The MRI-active zone is located inwards from the shaded area.
    }
    \label{fig:at}
\end{figure}

In the first part of our study, we perform 2D
locally isothermal simulations (Sect.~\ref{sec:model_iso}) with the viscosity transition
positioned roughly at $r_{\mathrm{MRI}}=1$ au. Locating the viscosity transition further out (with respect to 0.1 au)
means that a simulation covering tens of thousands of orbital timescales translates
to physical times that are comparable to the expected migration timescales in the Type II regime.
In other words, we are able to cover the migration history of gap-opening planets as they
approach the viscosity transition and as they interact with it.
Using the locally isothermal approximation, and thus neglecting the disk thermodynamics,
is not very realistic yet it simplifies the parametric space and it again helps to reduce
numerical demands. In the end, the simplicity can be advantageous because it can provide
hints on the robustness of our results.

Table~\ref{tab:params_varied} defines parameters that we vary in our simulations. They are specified throughout the following sections. Table~\ref{tab:params_iso},
on the other hand, summarizes parameters that are kept fixed. As we intend to use TOI-216b as our reference case, we adopt the mass of its host star $M_{\star}=0.77\,M_{\odot}$ and, unless specified differently, the planet-to-star mass ratio is
$q=7.3\times10^{-5}$.
The value of $h=0.023$ is motivated by the aspect ratio at the inner disk rim (see also Fig.~\ref{fig:noniso_disk}). As for the computational domain, it is chosen such that 
the Hill radius of TOI-216b is resolved by 7 cells and the half-width of the horseshoe region
\citep[e.g.][]{Jimenez_Masset_2017MNRAS.471.4917J} by 19 cells. The radial spacing is logarithmic
and the grid resolution is such that individual cells remain roughly square-shaped
to reduce numerical anisotropy.

\begin{figure*}[!t]
    \centering
    \begin{tabular}{cc}
        \includegraphics[width=0.95\columnwidth]{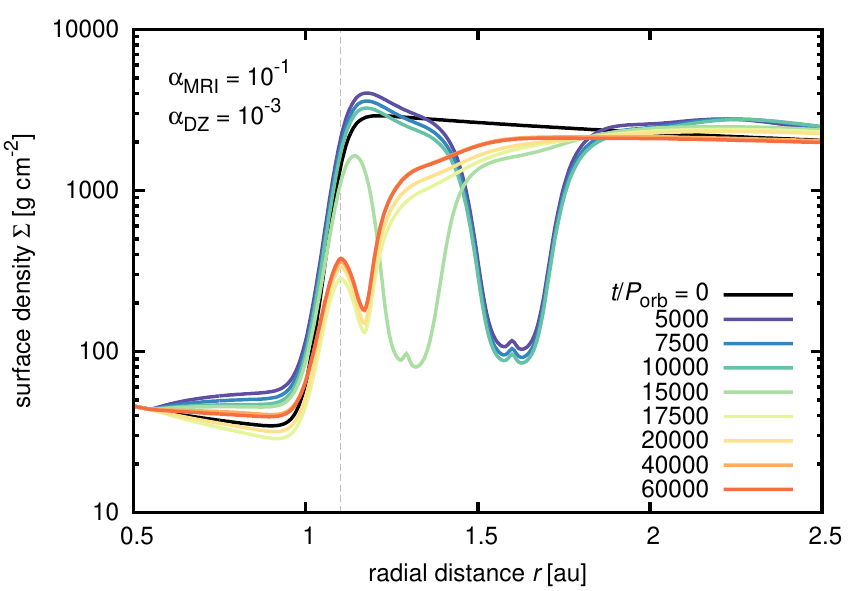} & \includegraphics[width=0.95\columnwidth]{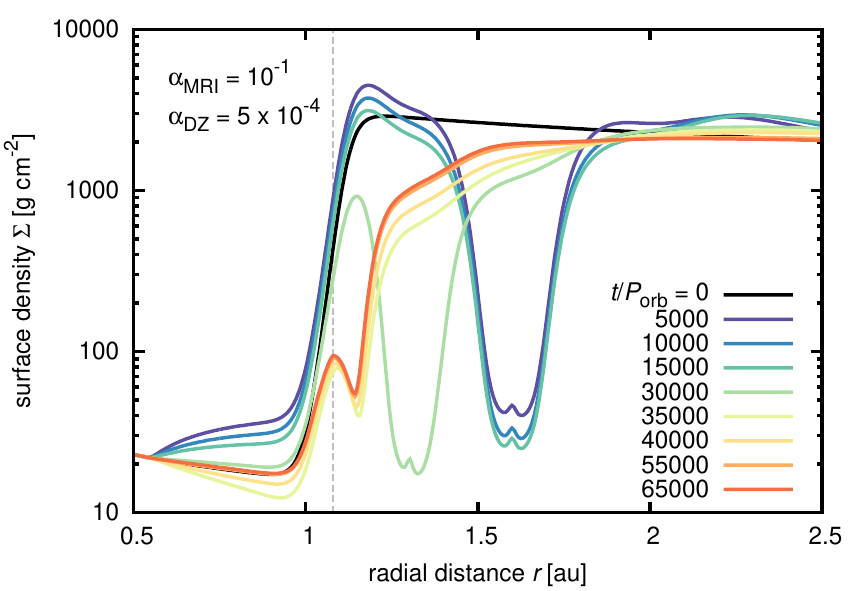} \\
        \includegraphics[width=0.95\columnwidth]{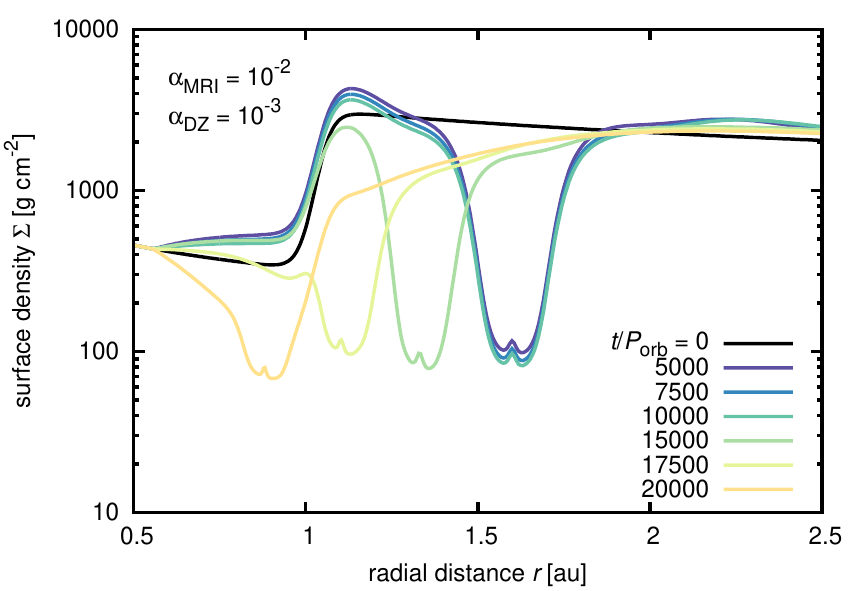} & 
        \includegraphics[width=0.95\columnwidth]{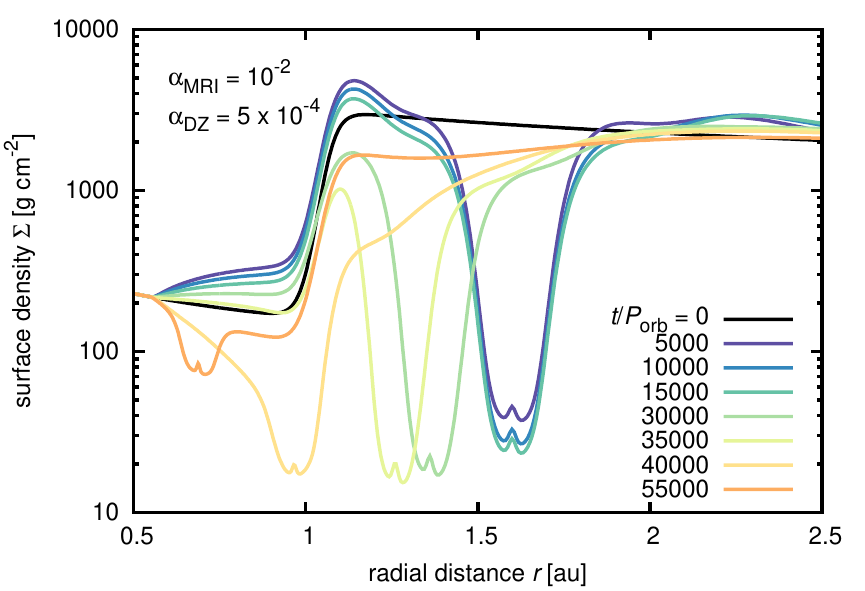}
    \end{tabular}
    \caption{Evolution of the radial surface density profile for four limit-case viscosity transitions.
    Simulations shown here combine $\alpha_{\mathrm{DZ}}=10^{-3}$ (left column) and $5\times10^{-4}$ (right column) with $\alpha_{\mathrm{MRI}}=10^{-1}$ (top row) and $10^{-2}$ (bottom row).
    The black curve corresponds to the unperturbed disk. Coloured curves correspond to different simulation
    times $t$ as indicated by the plot legend. The planet is kept fixed for $t=10^{4}$ and $1.5\times10^{4}\,\,P_{\mathrm{orb}}$ when $\alpha_{\mathrm{DZ}}=10^{-3}$ and $5\times10^{-4}$, respectively. At larger times, the planet is allowed to migrate freely. The vertical dashed line
    marks the final location of the planet in simulations where trapping occurs.
        We point out that by setting $\dot{M}$ based on $\alpha_{\mathrm{DZ}}$ (see Sect.~\ref{sec:migra_vs_alpha}),
    the unperturbed density profile in the dead zone is identical in each simulation.
    }
    \label{fig:sigma1d}
\end{figure*}

\subsection{The role of the $\alpha$-viscosity contrast for trapping TOI-216b}
\label{sec:migra_vs_alpha}

We start by placing TOI-216b at $a_{\mathrm{p}}(t_{0})=1.6\,\mathrm{au}$, outwards from the viscosity transition. The viscosity transition width parameter is set to $\Delta r_{\mathrm{MRI}}=2H=0.046\,\mathrm{au}$.
Two values are explored for the dead-zone viscosity, $\alpha_{\mathrm{DZ}}=5\times10^{-4}$ and $10^{-3}$.
For each of these $\alpha_{\mathrm{DZ}}$ values, we vary the MRI-active viscosity as
$\alpha_{\mathrm{MRI}}=(10^{-2},2.5\times10^{-2},5\times10^{-2},7.5\times10^{-2},10^{-1})$.
Since the disk-driven migration scales with the disk mass, it would be ideal to maintain the unperturbed
$\Sigma(r)$ profile the same in all our simulations. However, this can never be fulfilled
over the whole extent of the disk when varying the viscosities and initializing the disk via Eq.~(\ref{eq:mdot}) at the same time. We decided to maintain the same $\Sigma$ at least in the
dead zone of the disk, which is the region that (i) controls the approach of the planet towards
the viscosity transition and (ii) contains most of the disk mass in the given radial range. This is done by setting $\dot{M}=10^{-8}\,M_{\odot}\,\mathrm{yr}^{-1}$ when $\alpha_{\mathrm{DZ}}=10^{-3}$
and $\dot{M}=5\times10^{-9}\,M_{\odot}\,\mathrm{yr}^{-1}$ when $\alpha_{\mathrm{DZ}}=5\times10^{-4}$.
The planet mass is gradually increased over the first 1,000 orbital timescales and 
the planet itself is held in a fixed circular orbit for 10,000 and 15,000 orbital timescales when $\alpha_{\mathrm{DZ}}=10^{-3}$ and $5\times10^{-4}$, respectively. Only after a gap was formed
and its opening slowed down did we let the planet migrate for 50,000 $P_{\mathrm{orb}}$.

Since $a_{\mathrm{p}}(t_{0})=1.6\,\mathrm{au}$, our setup allows the gap to be first opened in the initially smooth part of the dead zone,
without any influence of the viscosity (and surface density) transition.
This way, the planet first acquires a correct migration rate before it approaches
the viscosity transition where the gap profile is expected to become modified.
We recall here that a gap-opening planet needs to be allowed to migrate
for some time before its migration rate converges \citep{Durmann_Kley_2015A&A...574A..52D,Robert_etal_2018A&A...617A..98R,Scardoni_etal_2020MNRAS.492.1318S,Kanagawa_Tanaka_2020MNRAS.494.3449K,Lega_etal_2021A&A...646A.166L}.
The reason is that the gap of a non-migrating planet represents a different barrier
(although leaky) for the background gas flow than a gap of a migrating planet \citep{Lubow_DAngelo_2006ApJ...641..526L,Duffell_etal_2014ApJ...792L..10D}.

Temporal evolution of the semi-major axis $a_{\mathrm{p}}$ after the release of TOI-261b
is shown in Fig.~\ref{fig:at} for all studied viscosity transitions. At first,
the migration proceeds as expected for a gap-opening planet. For a short time,
the planet experiences inward migration uncoupled from
the viscous flow in the disk due to the initial inbalance of the torques (as reflected by the slightly steeper slope at the very beginning of the migration curves). But then
the migration curves level at a slope tied to the viscous flow \citep[similarly to e.g.][]{Robert_etal_2018A&A...617A..98R}.
When the planet migrates across $a_{\mathrm{p}}=1.2$--$1.3\,\mathrm{au}$, the part of the inner spiral arm
that is propagating through the decreased density region below the viscosity transition
becomes progressively larger. Consequently, the outer spiral arm takes over and the Lindblad
torque becomes more negative, resulting in an episode of fast inward migration in the vicinity
of the viscosity transition.
What happens next depends on the viscosity values used to model the transition.
When $\alpha_{\mathrm{DZ}}=10^{-3}$, the planet is trapped at the viscosity transition
for $\alpha_{\mathrm{MRI}}\geq 5\times10^{-2}$, otherwise it continues migrating inwards.
When $\alpha_{\mathrm{DZ}}=5\times10^{-4}$, the trapping is only possible for $\alpha_{\mathrm{MRI}}\geq7.5\times10^{-2}$.
Such a dependence on the $\alpha$-viscosity suggests an interplay between the gap
depth and the trap efficiency---as the gap carved by the planet gets 
deeper in less viscous dead zones, larger
$\alpha_{\mathrm{MRI}}$ is then required to (partially) refill the gap so that the trapping mechanism can still work.
Therefore, we examine the gap profile next.

\begin{figure*}[!t]
    \centering
    \begin{tabular}{cc}
            (a) & (b) \\
        \includegraphics[width=0.95\columnwidth]{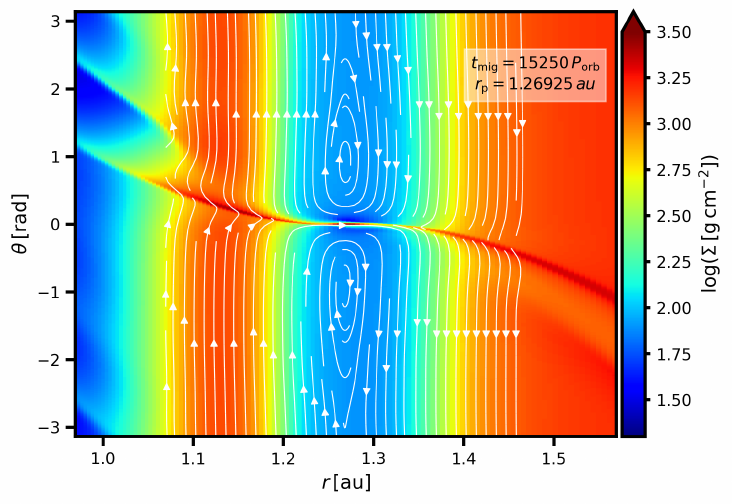} &
        \includegraphics[width=0.95\columnwidth]{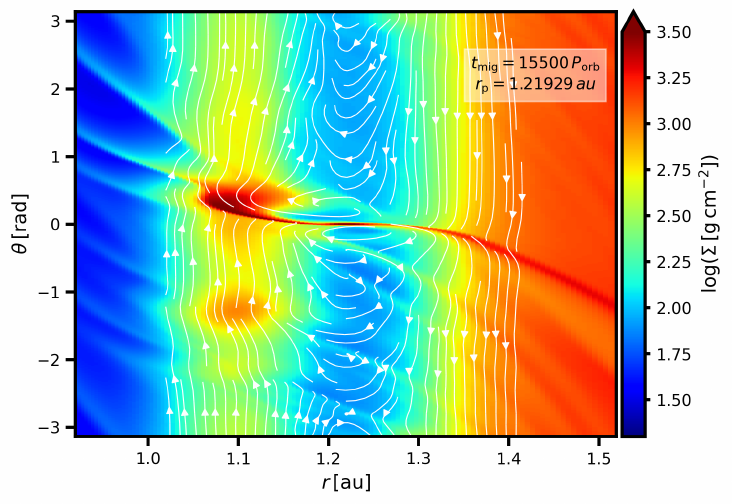} \\
            (c) & (d) \\
        \includegraphics[width=0.95\columnwidth]{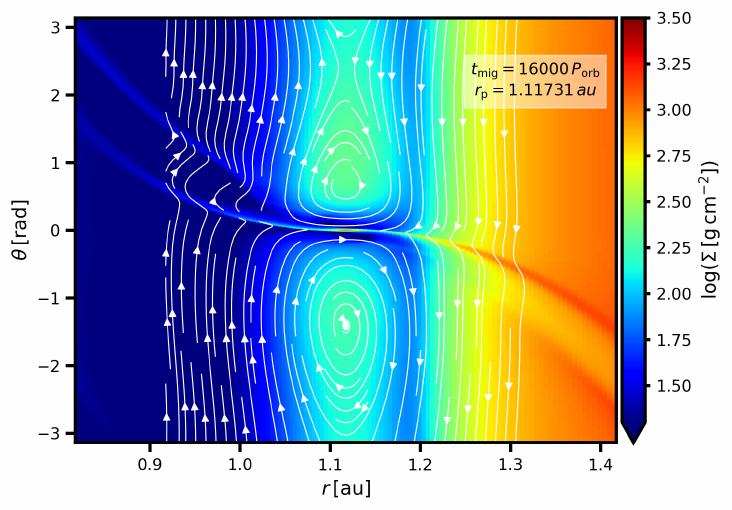} &
        \includegraphics[width=0.95\columnwidth]{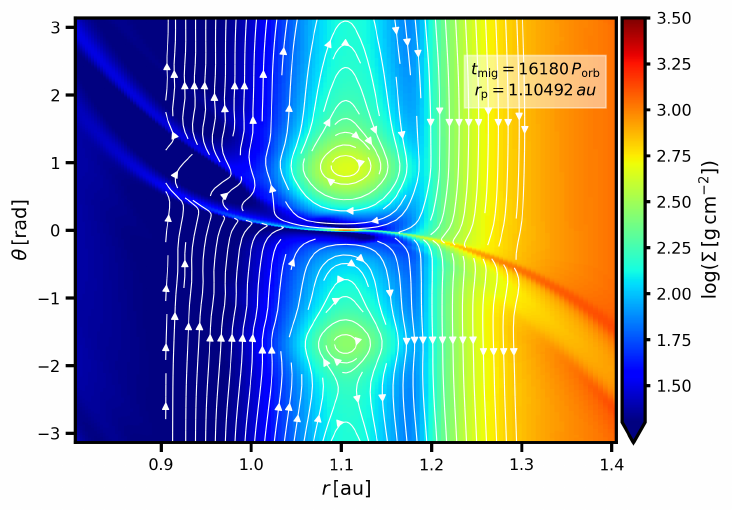} \\
            (e) & (f) \\
        \includegraphics[width=0.95\columnwidth]{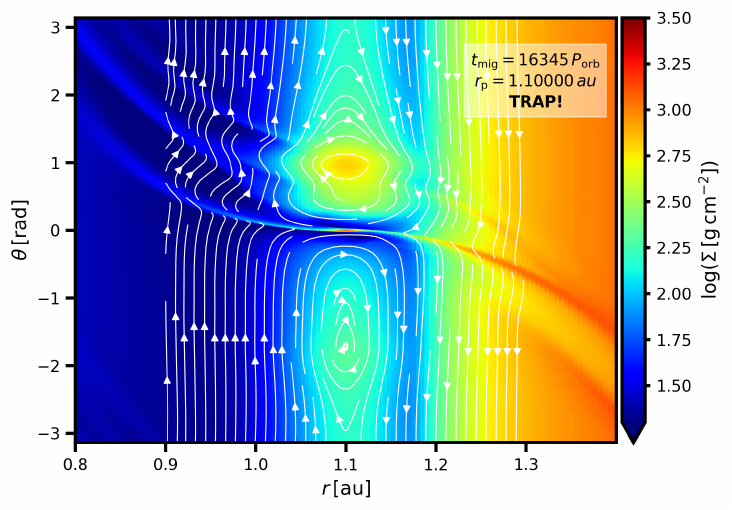} &
        \includegraphics[width=0.95\columnwidth]{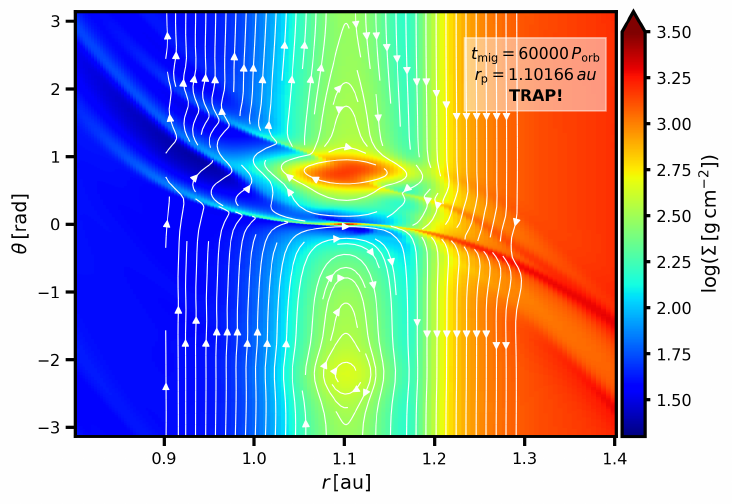}
    \end{tabular}
    \caption{Temporal evolution of the gas surface density $\Sigma$
    (shown in a logarithmic colour gradient) in a simulation with $\alpha_{\mathrm{DZ}}=10^{-3}$ and $\alpha_{\mathrm{MRI}}=10^{-1}$.
    In each panel, the depicted portion of the disk is centred on TOI-216b.
    A rectangular projection of azimuthal coordinates is used
    to highlight disk features orbiting at the same radial separation.
    Streamlines of the gas flow are overlaid as white oriented curves.
    Individual panels are labelled with the migration time $t_{\mathrm{mig}}$
    (corresponding to Fig.~\ref{fig:at}) and the instantaneous
    orbital radius of the planet $r_{\mathrm{p}}$. Panels (a)--(d)
    capture important phases of migration of TOI-216b towards the viscosity transition, panel (e) shows a state shortly after the migration stalls
    and panel (f) shows the final state of our simulation.
    The figure is also available as an online \texttt{movie} showing the temporal
    evolution between $t_{\mathrm{mig}}=16,000$ and $17,000\,P_{\mathrm{orb}}$.
    }
    \label{fig:fiduc_dens_evol}
\end{figure*}

\subsection{Evolution of the gap profile}
\label{sec:gap}

Fig.~\ref{fig:sigma1d} shows the temporal evolution of the radial surface density profile $\Sigma(r)$
for four selected simulations that combine the minimum and maximum $\alpha$-viscosities
from our parametric range. To obtain the $\Sigma(r)$ profiles, we calculated the arithmetic mean
from $\Sigma(r,\theta)$ in each radial ring of the domain. Fig.~\ref{fig:sigma1d} confirms
our predictions from \cite{Nesvorny_etal_2022ApJ...925...38N} and Sect.~\ref{sec:intro}---the gap
in the dead zone indeed becomes quite deep, with the minimum values
$\Sigma/\Sigma_{0}\simeq0.03$ when $\alpha_{\mathrm{DZ}}=10^{-3}$
and $\Sigma/\Sigma_{0}\simeq0.008$ when $\alpha_{\mathrm{DZ}}=5\times10^{-4}$.

We also immediately see a substantial difference in the gap profile between simulations
in which the planet is trapped and those in which it continues migrating inwards. For simulations where the trapping fails (bottom row in Fig.~\ref{fig:sigma1d}), we see that the gap mostly retains its characteristic shape even when the planet is crossing the viscosity transition---the gap centre still appears to be close to an inner-outer symmetry, with a small central
peak that reflects the mass accumulation in the Hill sphere of the planet. The gap walls do exhibit an inner-outer asymmetry, with the outer wall becoming steeper. As a result, the outer spiral arm
contains more gas mass relative to the inner spiral arm and thus drives a strong negative Lindblad torque that is responsible for the episodes of fast inward migration in Fig.~\ref{fig:at}.

When the trapping occurs (top row in Fig.~\ref{fig:sigma1d}), on the other hand, the gap profile becomes modified.
The inner half of the gap cannot be distinguished from the unperturbed density profile
and a local density peak occurs at the planet location. 
The outer half of the gap remains cleared of gas, although it is shallower compared
to simulations in which there is no trapping.
Due to the drop in the surface density across the viscous transition,
the outer spiral arm inevitably dominates over the inner one in this case as well.
We thus need to focus our attention on the changes in the co-rotation region
in order to understand why the torque felt by the planet becomes zero.
To do so, the next section analyses 2D maps of the surface density as well as gas flow properties.

\begin{figure}
    \centering
    \includegraphics[width=\columnwidth]{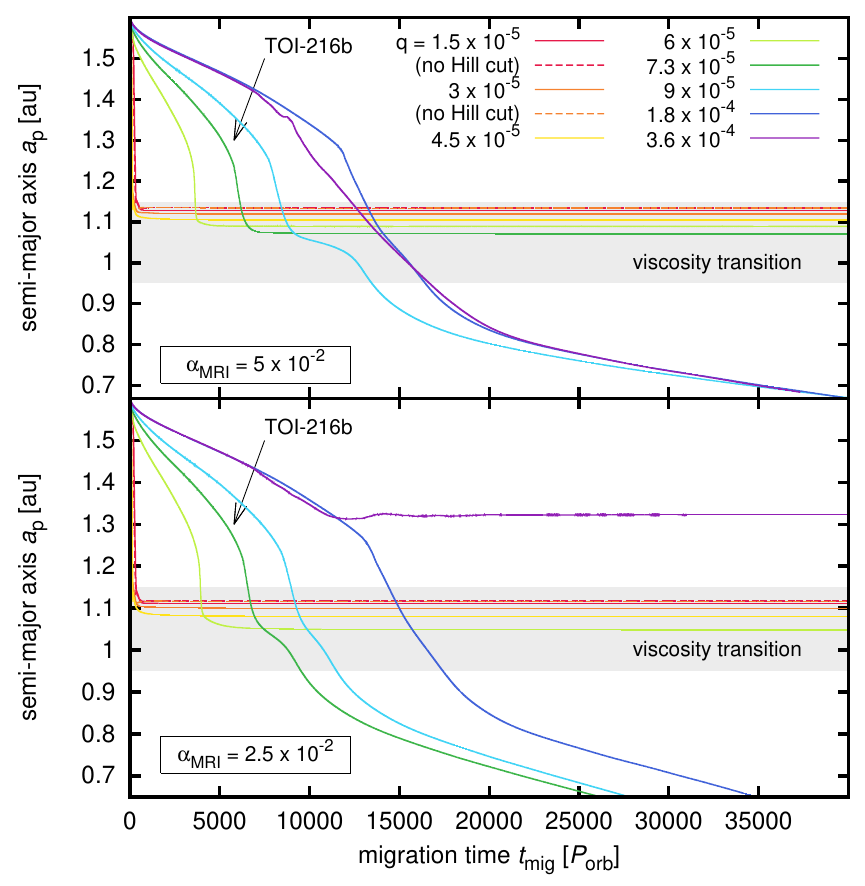}
    \caption{As in Fig.~\ref{fig:at}, but for fixed $\alpha_{\mathrm{DZ}}=10^{-3}$
    and different values of the planet-to-star mass ratio $q$ (see the plot legend).
    Two cases of the MRI-active viscosity are considered, namely $\alpha_{\mathrm{MRI}}=5\times10^{-2}$ (top)
    and $2.5\times10^{-2}$ (bottom). Dashed lines correspond to simulations in which the torque is evaluated
    from the whole disk, without any exclusion of the Hill sphere material. Solid lines are affected by the Hill
    cut (see Eq.~\ref{eq:hillcut}). For comparison purposes, the case of TOI-216b is labelled and marked with an arrow.
    }
    \label{fig:at_q}
\end{figure}

\begin{figure*}[!t]
    \centering
    \begin{tabular}{cc}
    (a) & (b) \\
        \includegraphics[width=0.95\columnwidth]{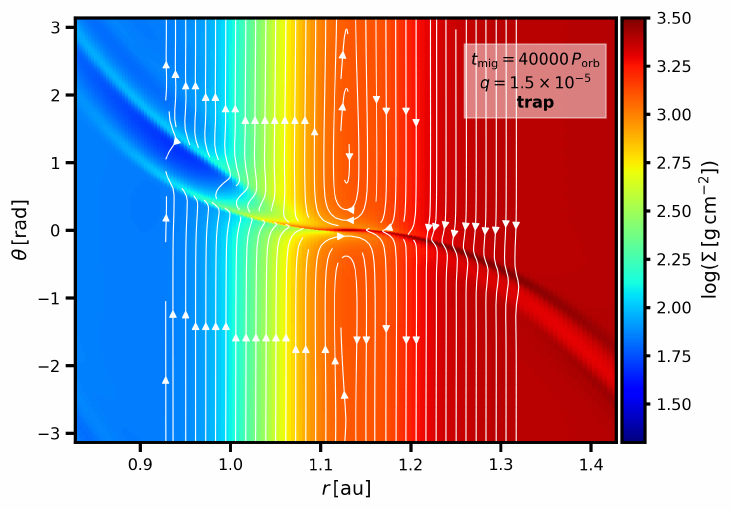} &
        \includegraphics[width=0.95\columnwidth]{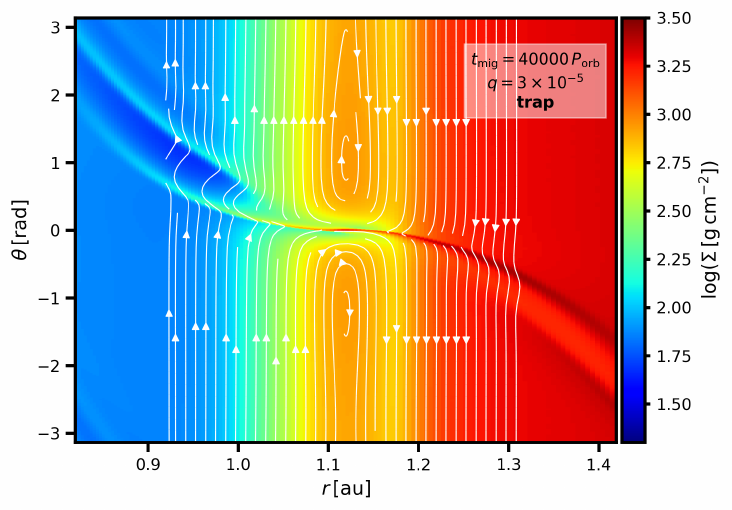} \\
        (c) & (d) \\
        \includegraphics[width=0.95\columnwidth]{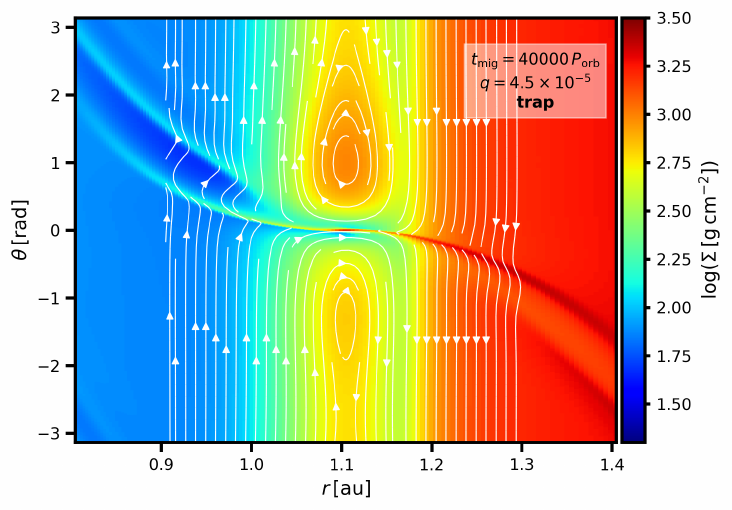} &
        \includegraphics[width=0.95\columnwidth]{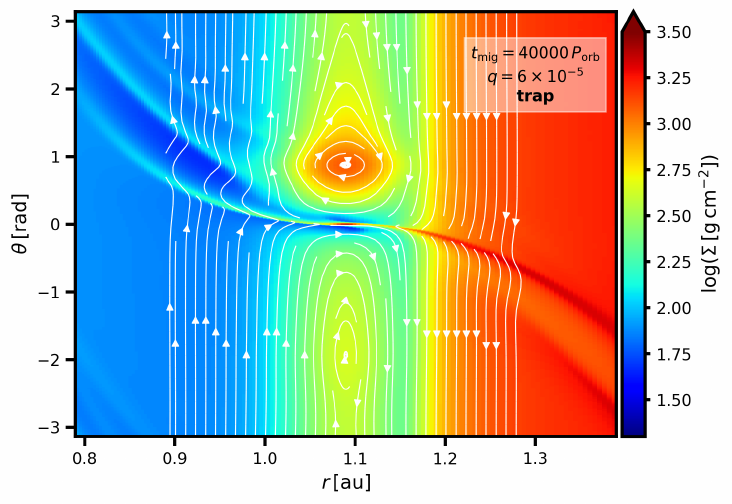} \\
        (e) & (f) \\
        \includegraphics[width=0.95\columnwidth]{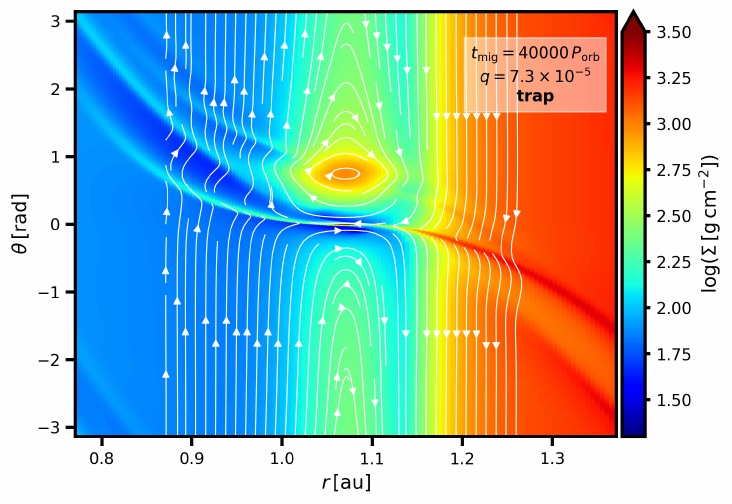} &
        \includegraphics[width=0.95\columnwidth]{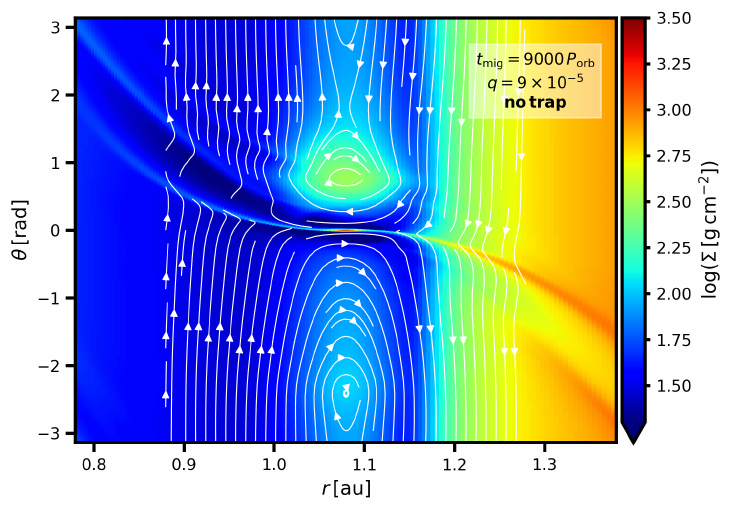}
    \end{tabular}
    \caption{As in Fig.~\ref{fig:fiduc_dens_evol}, but now for different
    values of the planet-to-star mass ratio $q$ as indicated by labels
    of individual panels. The viscosity transition is parameterized
    with $\alpha_{\mathrm{DZ}}=10^{-3}$ and $\alpha_{\mathrm{MRI}}=5\times10^{-2}$
    (see top panel of Fig.~\ref{fig:at_q} for corresponding 
    migration tracks of individual planets).
    In panels (a)--(e) that exhibit the migration trap,
    we show the state of the simulation at $t_{\mathrm{mig}}=40,000\,P_{\mathrm{orb}}$.
    In panel (f), the planet is not trapped and thus we show the state at
    $t_{\mathrm{mig}}=9,000\,P_{\mathrm{orb}}$
    when the planet is roughly at the same radial distance as in panel (e).
    }
    \label{fig:gasdens_q}
\end{figure*}

\begin{figure*}[!t]
    \centering
    \begin{tabular}{ccc}
        \includegraphics[width=0.64\columnwidth]{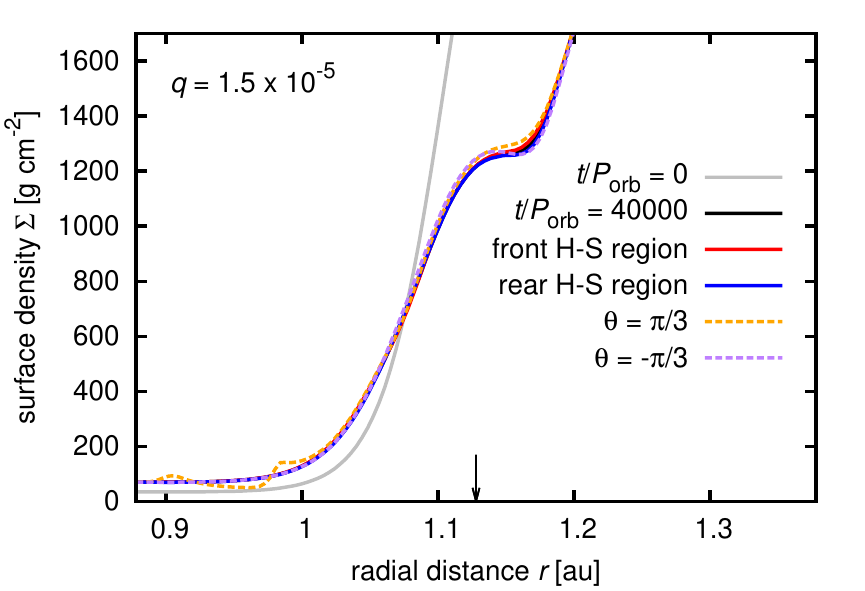} &
        \includegraphics[width=0.64\columnwidth]{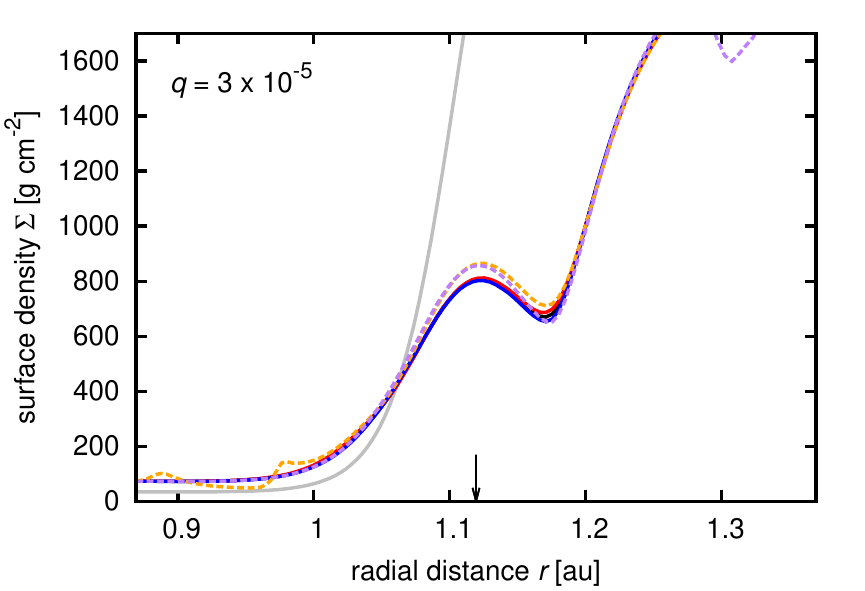} &
        \includegraphics[width=0.64\columnwidth]{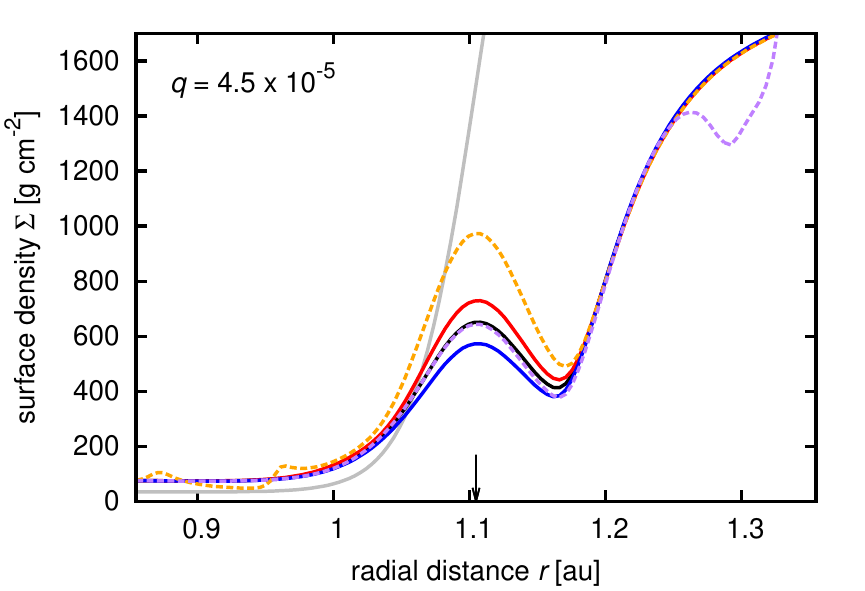} \\
        \includegraphics[width=0.64\columnwidth]{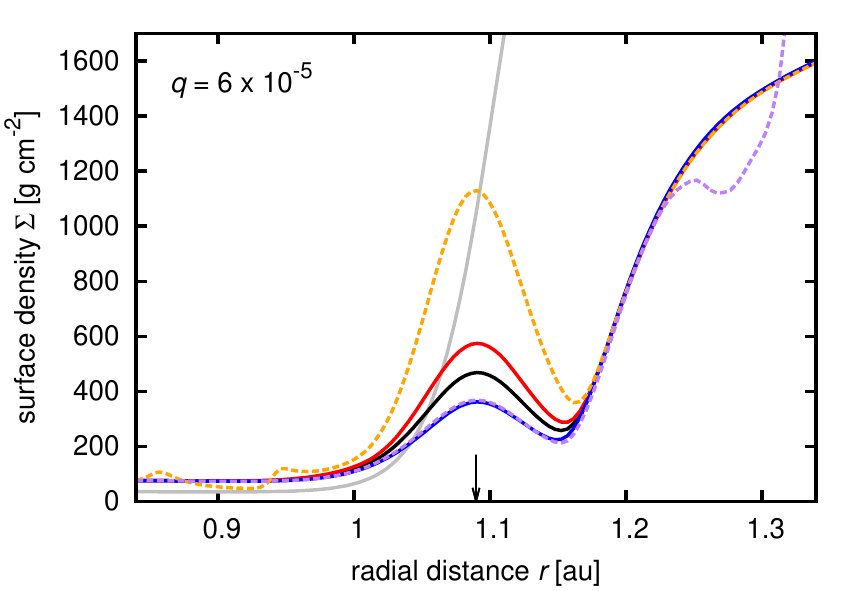} &
        \includegraphics[width=0.64\columnwidth]{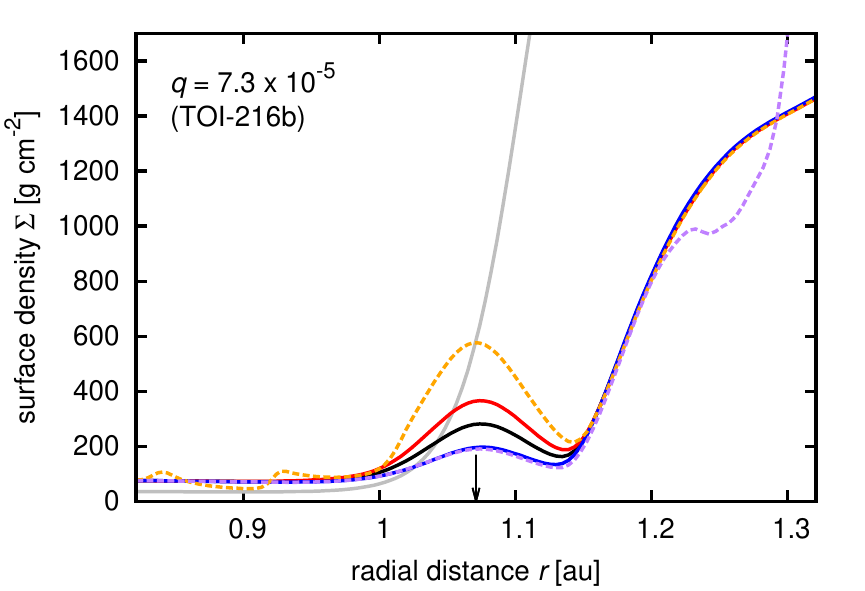} &
        \includegraphics[width=0.64\columnwidth]{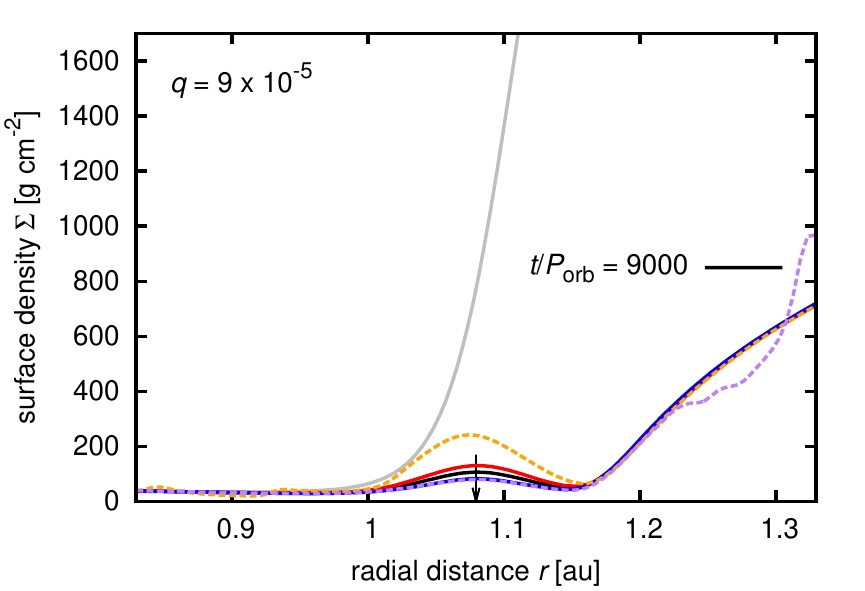}
    \end{tabular}
    \caption{Radial surface density profiles for cases discussed in Fig.~\ref{fig:gasdens_q}
    (planet-to-star mass ratios $q$ are given by labels in individual panels).
    We plot the initial azimuthal average (solid grey curve), the azimuthal average
    at $t_{\mathrm{orb}}=40,000\,P_{\mathrm{orb}}$ for trapped planets and at $t_{\mathrm{orb}}=9,000\,P_{\mathrm{orb}}$ for $q=9\times10^{-5}$ (black solid curve),
    azimuthal averages taken over the span of the front ($\theta>0$) and rear ($\theta<0$)
    horseshoe regions (red and blue solid curves, respectively),
    and radial cuts $\Sigma(r,\theta=\pi/3)$ and $\Sigma(r,\theta=-\pi/3)$ (orange and purple dashed curves, respectively).
    We point out that the Hill sphere material is excluded from calculations
    of all azimuthal averages.
    The instantaneous position of the planet is marked by the black arrow. Unlike in 
    Fig.~\ref{fig:sigma1d}, the vertical scale is linear.
    The spread of the curves (notably the orange and red ones) indicates the development
    of the front-rear asymmetry in the co-rotation region.
    }
    \label{fig:sigma1d_q}
\end{figure*}

\begin{figure}[!t]
    \centering
        \includegraphics[width=\columnwidth]{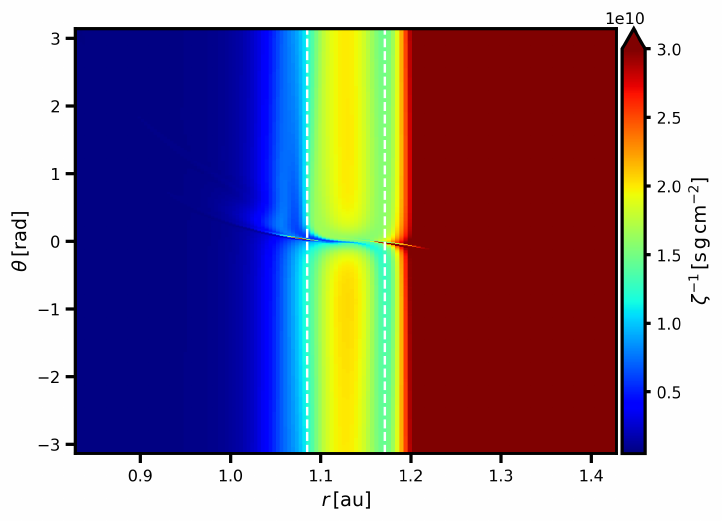}
        \includegraphics[width=\columnwidth]{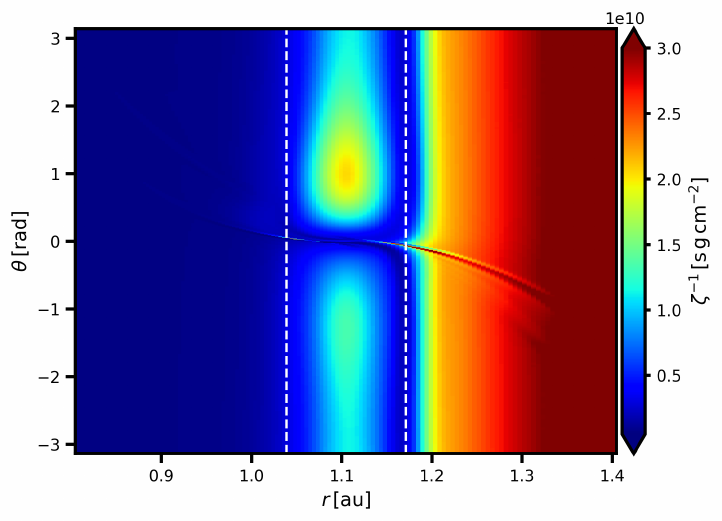}
        \includegraphics[width=\columnwidth]{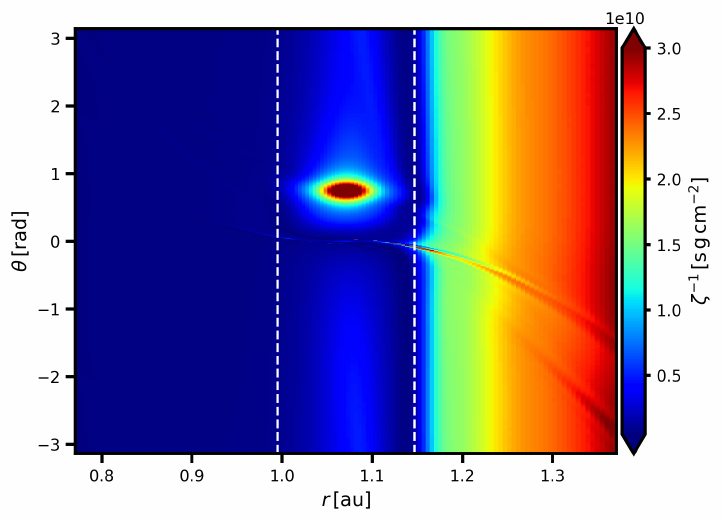}
    \caption{2D distribution of the inverse potential vorticity $\zeta^{-1}$
    for the same simulation set as in Figs.~\ref{fig:gasdens_q} and \ref{fig:sigma1d_q}
    and for $q=1.5\times10^{-5}$ (top), $4.5\times10^{-5}$ (middle) and $7.3\times10^{-5}$
    (bottom). Vertical dashed lines mark the approximate extent of the horseshoe region
    following \cite{Jimenez_Masset_2017MNRAS.471.4917J} (their Eq. 30).
    The colour scale is saturated to highlight features in the co-rotation region.}
    \label{fig:invPV_q}
\end{figure}

\subsection{Surface density and gas flow evolution in the co-rotation region}
\label{sec:dynamic_vortex}

Fig.~\ref{fig:fiduc_dens_evol} shows several snapshots from our hydrodynamic
simulation of migrating TOI-216b that becomes trapped at the viscous
transition between $\alpha_{\mathrm{DZ}}=10^{-3}$ and $\alpha_{\mathrm{MRI}}=10^{-1}$.
Panel (a) shows TOI-216b while it is still positioned outside the outer edge of the viscosity
transition. The gap is regular as well as the streamlines within.
However, as the planet continues approaching
the viscosity transition, the gas density distribution at the inner gap edge 
turns into a radially narrow bump. This can be seen near $\simeq$$1.1\,\mathrm{au}$ in panel (b)
where the initial step-like profile of $\Sigma(r)$ turns into a bump
that is bordered by the low-density high-viscosity region on the inside and by the planet-induced gap on the outside.
The density bump becomes prone to the Rossby wave 
instability \citep[RWI;][]{Lovelace_etal_1999ApJ...513..805L,Li_etal_2000ApJ...533.1023L} that manifests itself via a high-$m$ mode and excites multiple vortices.

The instability does not persist as the gap follows the inward-migrating planet
and reduces the gas accumulation at $\simeq$$1.1\,\mathrm{au}$ even more, as shown
in panel (c). Nevertheless, a less prominent local density bump is still maintained
within the co-rotation region of the planet. The streamlines in the co-rotation region
develop two librating islands, one positioned in the front horseshoe region
(leading the orbital motion of the planet) and one located in the rear horseshoe
region (trailing the orbital motion of the planet). We notice a front-rear asymmetry,
with the front island being azimuthally narrower.
Over time, both islands of librating streamlines accumulate gas but the front one 
is more efficient as indicated by panel (d).

Eventually, the co-rotation streamlines reconfigure when the rear overdensity 
propagates through the co-rotation region in a retrograde sense (with respect to the planet)
and merges with the front overdensity. The state at the time of the merger is depicted
in panel (e). At this $t_{\mathrm{mig}}$, the migration trap is first detected ($\mathrm{d}a_{\mathrm{p}}/\mathrm{d}t$ becomes non-negative for the first time).
The overdensity in the front horseshoe region becomes a Rossby vortex and launches its
own set of spiral arms. The final state of the simulation is shown in panel (f).
The vortex in the front libration island persists and its peak density (with respect
to panel (e)) increases.

Overall, it is clear that the trapping is facilitated by the co-rotation torque
but its nature is different compared to \cite{Masset_etal_2006ApJ...642..478M}.
According to \cite{Masset_etal_2006ApJ...642..478M}, planets
stop because the strong vortensity gradient at the surface density transition
modifies the horseshoe drag \citep{Ward_1991LPI....22.1463W}.
Here, instead, the partial gap opening reduces the initial gas density and changes its
shape from a step-like to a bump-like. Thus the vortensity gradient is diminished
as well as the actual amount of gas in the co-rotation region that can exert the torque.
The trapping is only possible due 
to the front-rear asymmetry that occurs in the co-rotation region.
The front overdensity dominates and exerts a positive torque on the planet
(as any azimuthal overdensity leading the orbital motion of the planet would)
that balances the negative torque of spiral arms. Consequently, we find that
the migration stops.

Although the nature of the trap for TOI-216b is now clear,
it is not a priori obvious what is the main driving mechanism for the
occurrence of islands of librating streamlines. On one hand, we detect vortices in panels
(b), (e) and (f) of Fig.~\ref{fig:fiduc_dens_evol} and those would suggest that the RWI
dominates. On the other hand, panels (c) and (d) already exhibit the libration islands
but there is no apparent sign of vortices and that would suggest that the islands 
are a natural feature of a partially emptied co-rotation region that overlaps
a viscosity transition. To gain further insight concerning this problem, 
we explore the behaviour of the planet trap for various planet-to-star
mass ratios $q$ in the upcoming section.

\subsection{Varying the planetary mass}
\label{sec:vary_mass}

In previous sections, the planetary mass was fixed to that of TOI-216b. In this section,
we investigate the planet-to-star mass ratios\footnote{Our range of $q$ would translate to planetary masses $M_{\mathrm{p}}=(3.85,7.7,11.55,15.4,23.1,46.2,92.4)\,M_{\oplus}$ in
the TOI-216 system or to $M_{\mathrm{p}}=(5,10,15,20,30,60,120)\,M_{\oplus}$ in the Solar
System.} 
$q=(1.5,3,4.5,6,9,18,36)\times10^{-5}$. Fig.~\ref{fig:at_q} shows the temporal evolution of the planetary semi-major axes $a_{\mathrm{p}}(t)$
for fixed $\alpha_{\mathrm{DZ}}=10^{-3}$ and two values of the MRI-active viscosity $\alpha_{\mathrm{MRI}}=2.5\times10^{-2}$ and $\alpha_{\mathrm{MRI}}=5\times10^{-2}$. We selected these two viscosity values because
they represent the divide between trapping and not trapping TOI-216b (see the labelled green curve in Fig.~\ref{fig:at_q}).
The remaining parameters remain the same as in Sect.~\ref{sec:migra_vs_alpha}
apart from the total simulation timescale after the release of the planet,
which we reduced to $40,000\,P_{\mathrm{orb}}$.
In terms of the trapping efficiency, Fig.~\ref{fig:at_q} reveals that all studied planets less massive than
TOI-216b become trapped at the viscosity transition while more massive planets avoid the trap and continue
migrating inwards (with one peculiar exception mentioned below).

It is also worth noting that planets with $q\leq4.5\times10^{-5}$ experience a runaway migration over the time
interval $100$--$350\,P_{\mathrm{orb}}$ that is barely resolved in Fig.~\ref{fig:at_q}. To exclude the possibility
that the runaway migration occurs due to the exclusion of some of the gas in the Hill sphere from the torque evaluation, which might not be fully justified for low values of $q$, we recalculated the two least massive
cases while considering the torque exerted by the whole disk (dashed lines in Fig.~\ref{fig:at_q}). The Hill cut
has a minimal influence on the semi-major axis evolution. The reason for the
fast inward migration is that these planets open partial gaps and then they enter the Type III migration regime \citep{Masset_Papaloizou_2003ApJ...588..494M,Peplinski_etal_2008MNRAS.386..164P}.
To our knowledge, this is the first numerical suggestion that super-Earths or mini-Neptunes might undergo inward Type III migration
under the conditions specific for the inner disk region (mainly the very low aspect ratio $h$ and 
a suitably large $\Sigma$). Despite their runaway migration, these super-Earths or mini-Neptunes become trapped at the
viscosity transition efficiently.
We refer the reader to Appendix~\ref{sec:app_add} and Fig.~\ref{fig:peculiar1} for an extended discussion.

The most massive planet in the bottom panel of Fig.~\ref{fig:at_q} (purple curve)
stops migrating before reaching the viscosity transition
because it develops a significant eccentricity equal to $e=0.05$
and then the torque becomes modified by interactions with the gap walls \citep{Duffell_Chiang_2015ApJ...812...94D}.
However, it is possible that this simulation suffers from numerical artefacts and an extended study
would be needed to exclude or confirm such a possibility. The reason for doubt is that the dead zone
properties are identical for both panels of Fig.~\ref{fig:at_q} yet the planet behaves differently in the bottom 
panel. We provide additional explanatory points in Appendix~\ref{sec:app_add} and Fig.~\ref{fig:peculiar2}.

In Fig.~\ref{fig:gasdens_q}, we show 2D maps of the gas density and streamlines
in the vicinity of each considered planet for $\alpha_{\mathrm{DZ}}=10^{-3}$ and $\alpha_{\mathrm{MRI}}=5\times10^{-2}$. With the increasing planetary mass,
the gap-opening capability becomes more efficient and we see the following trends: (i) the
surface density transition at the inner rim becomes gradually
erased and replaced by a density bump centred in the co-rotation region;
(ii) the density 
in the co-rotation region and thus the total mass of gas that can exert the co-rotation
torque decreases on average (see also Fig.~\ref{fig:sigma1d_q}); (iii) the streamline topology
changes from regular horseshoes towards a dominant front libration island; (iv) a front-rear
asymmetry in the gas density of the co-rotation region develops; (v) spiral arms launched
by a front vortex are apparent for $q\geq6\times10^{-5}$.

In Fig.~\ref{fig:gasdens_q}, planets with $q\leq7.3\times10^{-5}$ are trapped.
Based on this fact and based on the description provided in the previous paragraph,
we conclude that the discussed simulation set captures a transition from the classical trap
of \cite{Masset_etal_2006ApJ...642..478M} (that should dominate e.g. for $q=1.5\times10^{-5}$ for which we do not see any strong front-rear asymmetry in the gas distribution)
to a trap driven by the front-rear asymmetry of the co-rotation region for planets with moderate gaps (e.g. $q=4.5\times10^{-5}$--$7.3\times10^{-5}$) 
and finally to untrappable planets that clear their co-rotation regions too efficiently (e.g. $q=9\times10^{-5}$).

To further support these claims, Fig.~\ref{fig:sigma1d_q} shows the radial surface density
profiles for the cases discussed in Fig.~\ref{fig:gasdens_q}. We can see in detail how the initial step-like transition of $\Sigma(r)$ changes to a 
bump-dominated profile, whose amplitude decreases with increasing $q$.
Additionally, by plotting the azimuthal averages over front and rear horseshoe regions
and also radial cuts through the domain at $\theta\pm\pi/3$, we reveal the development
and magnitude of the front-rear asymmetry of the horseshoe region.
The most prominent asymmetry occurs in the range $q=4.5$--$7.3\times10^{-5}$
and for these planets, it produces the positive torque necessary for their trapping.
For $q=9\times10^{-5}$, the failure of the migration trap is caused by the decreased
level of the front-rear asymmetry and also by the overall emptying of the co-rotation region.

Finally, let us return to the question of the relative importance of the RWI compared to
planet-induced perturbations of the gas flow (see the end of Sect.~\ref{sec:dynamic_vortex}). To proceed with the analysis,
Fig.~\ref{fig:invPV_q} shows maps of the inverse potential vorticity
(or inverse vortensity) defined as
\begin{equation}
    \zeta^{-1} = \frac{\Sigma}{\left(\nabla\times\vec{v}\right)_{\perp} + 2\Omega_{\mathrm{p}}} \, ,
\end{equation}
where the vorticity component is perpendicular to the disk plane and the term related to
the orbital frequency translates $\zeta^{-1}$ to the inertial frame.
The inverse potential vorticity has several diagnostic applications in our case.
First, we look for the local maxima of $\zeta^{-1}$ to trace
vortices that are the usual outcomes of non-linear RWI growth\footnote{We emphasize
that we search for the local extrema of $\zeta^{-1}$ merely to localize the vortices
but not to assess the disk stability to the RWI. The latter cannot be done straightforwardly
because, as pointed out by e.g. \cite{Hallam_Paardekooper_2020MNRAS.491.5759H}, 
one should use the entropy-modified inverse potential vorticity as the key function
for locally isothermal disks (because they are not barotropic) but even then the extremum
of the key function does not guarantee the growth of the RWI, the key function would also need to exceed an a priori unknown threshold \citep{Li_etal_2000ApJ...533.1023L}. For these reasons, we use $\zeta^{-1}$ that should at least trace the vortices and from those we can conclude that the RWI is ongoing.}.
Second, the vortensity-driven horseshoe drag scales as $\Gamma_{\mathrm{HS}}\propto\Sigma\mathrm{d}\log \zeta^{-1}/\mathrm{d}\log r$ \citep{Masset_etal_2006ApJ...642..478M}
so that steep radial gradients of $\zeta^{-1}$ can help us judge 
if the classical migration trap of \cite{Masset_etal_2006ApJ...642..478M} still applies or not.

In Fig.~\ref{fig:invPV_q}, we focus on three cases $q=(1.5,4.5,7.3)\times10^{-5}$.
For the lowest planet mass, we see that a steep positive gradient of $\zeta^{-1}$
across the co-rotation is maintained and that confirms that the planet
is trapped primarily due to the mechanism of \cite{Masset_etal_2006ApJ...642..478M}.
Additionally, we notice a small asymmetry at the outer separatrix of the horseshoe
region---its front part has a slightly larger $\zeta^{-1}$ compared to 
the rear part. This is due to two effects: (i) the vorticity production
across the shock front of the outer spiral arm \citep{Koller_etal_2003ApJ...596L..91K}
that propagates through a low-viscosity dead zone where the ability
of the planet to shock the disk is higher \citep{Rafikov_2002ApJ...572..566R}
and (ii) the low-density material being carried from the inner disk towards the outer
disk by rear horseshoes \citep{Romanova_etal_2019MNRAS.485.2666R}. 
Both effects (growing $\nabla\times\vec{v}$ and mixing with a lower $\Sigma$)
reduce $\zeta^{-1}$ along the outer separatrix of the rear horseshoe region.

The middle and bottom panels of Fig.~\ref{fig:invPV_q} show that the gradient of
$\zeta^{-1}$ across the co-rotation is reduced with increasing planet mass, as
we already anticipated. Moreover, we notice that a steep isolated local maximum of
$\zeta^{-1}$ only appears for $q=7.3\times10^{-5}$. For $q=4.5\times10^{-5}$,
the peak values do not exceed those found for $q=1.5\times10^{-5}$ and
thus the planet with $q=4.5\times10^{-5}$ does not develop a vortex in its co-rotation
region (which is consistent with the lack of vortex-driven spiral arms for this planet mass).
Yet the co-rotation region of $q=4.5\times10^{-5}$ already exhibits a front-rear asymmetry dominated by the front island of librating streamlines. In addition, we recall that the azimuthally averaged density bump
across the co-rotation is actually larger for $q=4.5\times10^{-5}$ than for $q=7.3\times10^{-5}$ (panels (c) and (e), respectively, in Fig.~\ref{fig:sigma1d_q}).
The fact that the sharper peak in gas density remains Rossby-stable
means that the vortex found for $q=7.3\times10^{-5}$ has to be a secondary effect
launched by the streamline perturbations due to the influence of the planet.
This answers the question on the relative importance of these two effects:
the islands of librating streamlines develop regardless of the RWI
but the RWI can be triggered secondarily if an island accumulates enough mass or exhibits
a substantial change in the curl of the gas velocity.

Before moving on, let us point out that although we studied the dependence on
$q$ in this section, the described effects are related rather to the 
gap-opening ability of a planet that scales according to Eqs.~(\ref{eq:kanagawa}) and (\ref{eq:K}). Therefore, similar effects can be achieved by variations of $\alpha_{\mathrm{DZ}}$ and $\alpha_{\mathrm{MRI}}$ (which is consistent with results reported in Sect.~\ref{sec:migra_vs_alpha})
or by varying $h$ (not explored in this paper).

\subsection{Varying the viscosity transition width}
\label{sec:vary_width}

\begin{figure}
    \centering
    \includegraphics[width=\columnwidth]{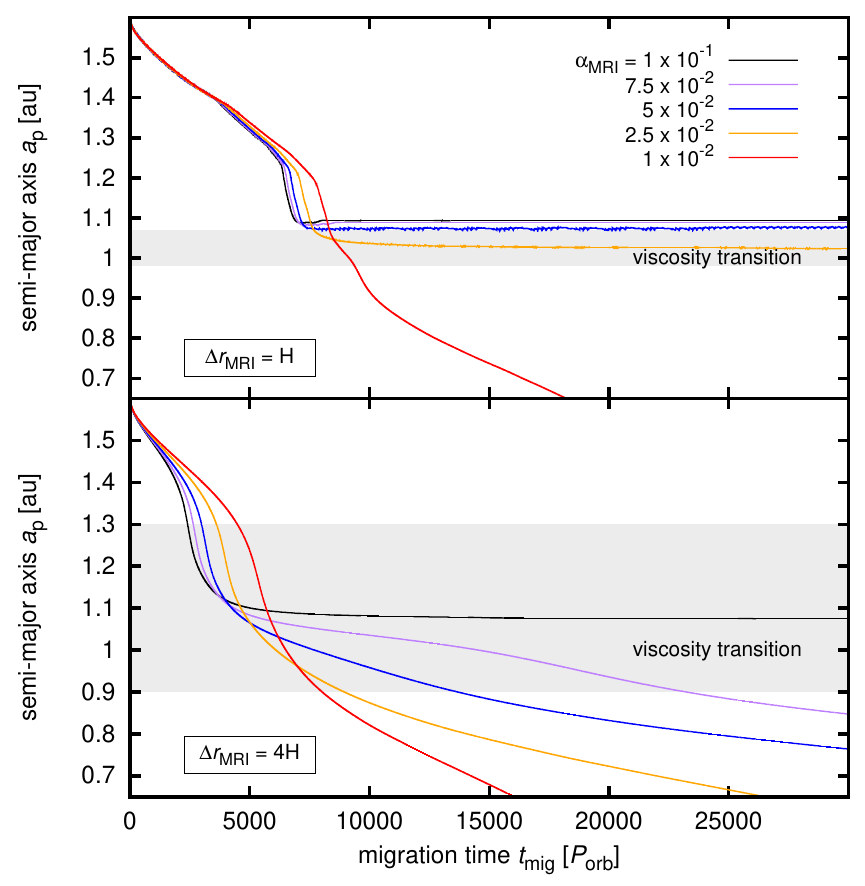}
    \caption{As in Fig.~\ref{fig:at}, but for fixed $\alpha_{\mathrm{DZ}}=10^{-3}$
    and different values of the viscosity transition width parameter $\Delta r_{\mathrm{MRI}}=H=0.023\,\mathrm{au}$ (top) and $\Delta r_{\mathrm{MRI}}=4H=0.092\,\mathrm{au}$ (bottom).
    We point out that the total width of the viscosity transition (as marked by the shaded area)
    is not exactly equal to $\Delta r_{\mathrm{MRI}}$; the influence of the parameter 
    is described by Eq.~(\ref{eq:transition}).
    }
    \label{fig:at_dr}
\end{figure}

In this section we modify the parameter $\Delta r_{\mathrm{MRI}}$ that controls
the width of the viscosity transition. We set $\alpha_{\mathrm{DZ}}=10^{-3}$, $\alpha_{\mathrm{MRI}}=(10^{-2},2.5\times10^{-2},5\times10^{-2},7.5\times10^{-2},10^{-1})$
and we explore two cases with $\Delta r_{\mathrm{MRI}}=H=0.023\,\mathrm{au}$ and $\Delta r_{\mathrm{MRI}}=4H=0.092\,\mathrm{au}$. The remaining parameters follow Sect.~\ref{sec:migra_vs_alpha} with the exception of the total simulation timescale after the release of the planet, which we reduced to $30,000\,P_{\mathrm{orb}}$.

The migration tracks that we obtained are shown in Fig.~\ref{fig:at_dr}. We see that if the viscosity transition
is sudden, the range of viscosities in the MRI-active zone that facilitate the trapping increases to $\alpha_{\mathrm{MRI}}\geq2.5\times10^{-2}$ compared to Fig.~\ref{fig:at}. 
If, on the other hand, the viscosity transition becomes flattened, the trapping becomes
increasingly difficult and only occurs for the most turbulent MRI zone with $\alpha_{\mathrm{MRI}}=10^{-1}$.

Our exploration of the parametric space does not go beyond the simple experiment shown 
in Fig.~\ref{fig:at_dr} yet it reveals that the width of the viscosity transition might be equally
important for the efficiency of the migration trap as the actual values of the viscosity itself.
For a reference, we recall that \cite{Flock_etal_2017ApJ...835..230F} found rather sharp viscosity transitions occurring over 1 or 2H but their Ohmic resistivity model was simplified. 
Future magnetohydrodynamic studies should therefore improve the ionization profile
and disk chemistry to carefully reassess the radial disk range over which the MRI is triggered.

\subsection{Exploring the static torque}
\label{sec:static}

\begin{figure}[!t]
    \centering
    \includegraphics[width=\columnwidth]{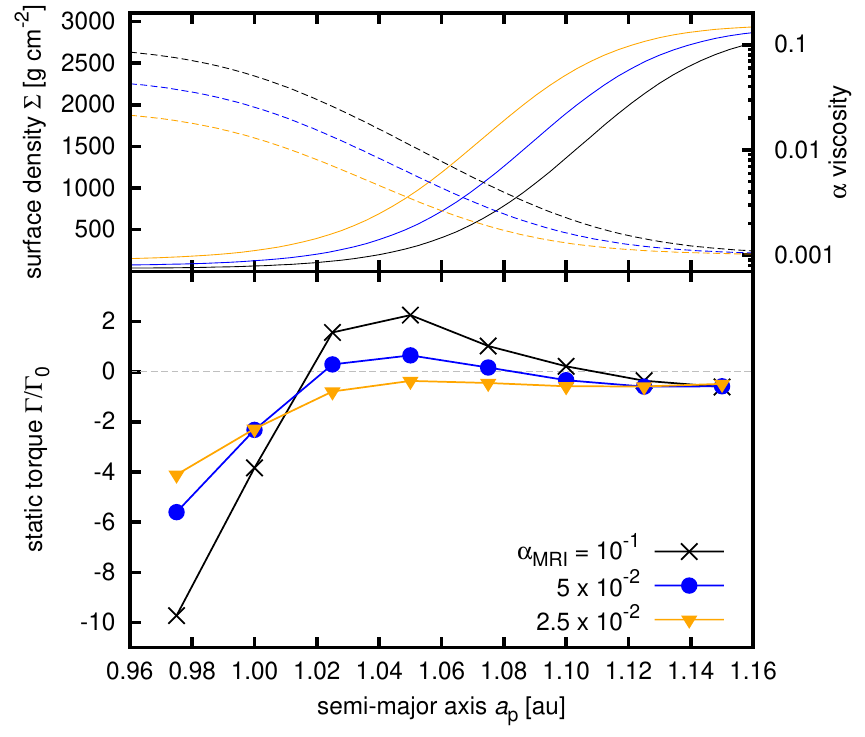}
    \caption{Locally isothermal simulations with a non-migrating planet. Top: Unperturbed surface density profile (solid lines, left vertical axis) across
    the viscosity transition (dashed lines, right vertical axis). Bottom:
    Measurements of the normalized static torque at various orbital radii (points)
    scanning the viscosity transition.
    In both panels, $\alpha_{\mathrm{DZ}}=10^{-3}$ and different values of $\alpha_{\mathrm{MRI}}$
    are distinguished by colours and symbols (see the plot legend). 
    }
    \label{fig:statictq_iso}
\end{figure}

\begin{figure}[!t]
    \centering
        \includegraphics[width=\columnwidth]{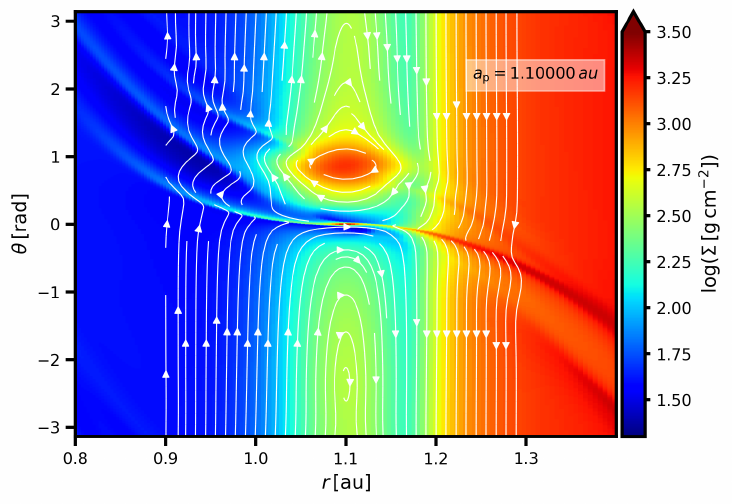}
    \caption{Surface density and streamlines for viscosities
    $\alpha_{\mathrm{DZ}}=10^{-3}$, $\alpha_{\mathrm{MRI}}=10^{-1}$
    and TOI-216b in a fixed circular orbit at $a_{\mathrm{p}}=1.1\,\mathrm{au}$.
    The plot serves as a comparison with panel (f) of Fig.~\ref{fig:fiduc_dens_evol}
    where the planet was migrating.
    The figure is also available as an online \texttt{movie} showing the
    variations in $\Sigma$ and streamlines for all values of $a_{\mathrm{p}}$
    considered in our static torque measurements.
    }
    \label{fig:statictq_dens}
\end{figure}

In previous sections, we saw that the migration trap for sub-Neptunes and Neptunes
at the inner rim is caused by the front-rear asymmetry of the co-rotation region.
Similar asymmetries, although of a different origin, often occur as a result 
of planet migration that leads to streamline deformation and, since they would not occur for planets in fixed orbits, the resulting torques are usually referred to
as dynamical torques.
Here, we perform simulations of TOI-216b in a fixed circular orbit and measure the static torque (i.e. the torque acting on a non-migrating planet)
for a set of semi-major axes $a_{\mathrm{p}}$
distributed in a close proximity of the viscosity transition. Our aim is to check whether the static torque measurements can also recover the trap and its position compared
to simulations with a migrating planet (Sect.~\ref{sec:migra_vs_alpha}). In other words, we want
to check whether the trapping mechanism depends on the migration history of the planet or not, which is an important question.

The parameters remain the same as in Sect.~\ref{sec:migra_vs_alpha}
but we limit the explored viscosities to $\alpha_{\mathrm{DZ}}=10^{-3}$ and $\alpha_{\mathrm{MRI}}=(2.5\times10^{-2},5\times10^{-2},10^{-1})$. Additionally,
the planetary semi-major axis remains fixed and we take $a_{\mathrm{p}}$ distributed evenly between $0.975\,\mathrm{au}$ and $1.15\,\mathrm{au}$ with a step of $\Delta a_{\mathrm{p}}=0.025\,\mathrm{au}$. The planetary 
mass is again increased over the first $1,000\,P_{\mathrm{orb}}$ and each simulation
covers $20,000\,P_{\mathrm{orb}}$ to allow the disk torque to converge.

Fig.~\ref{fig:statictq_iso} shows our measurements of the static torque $\Gamma$ normalized
to $\Gamma_{0}=\Sigma_{\mathrm{p}}a_{\mathrm{p}}^{4}(\Omega_{\mathrm{p}}q/h_{\mathrm{p}})^{2}$ \citep[e.g.][]{Korycansky_Pollack_1993Icar..102..150K},
where the subscript refers to the planet location and the surface density is that of 
an unperturbed disk. Wherever $\Gamma/\Gamma_{0}>0$ in Fig.~\ref{fig:statictq_iso},
outward migration is expected and vice versa. For $\alpha_{\mathrm{MRI}}=10^{-1}$, 
linear interpolation between the static torque measurements predicts trapping at $a_{\mathrm{p}}=1.1095\,\mathrm{au}$
compared to $1.1009\,\mathrm{au}$ found for the migrating planet.
For $\alpha_{\mathrm{MRI}}=5\times10^{-2}$, the zero-torque radius is located at $a_{\mathrm{p}}=1.0834\,\mathrm{au}$
compared to the dynamical trap location $1.0708\,\mathrm{au}$.
For $\alpha_{\mathrm{MRI}}=2.5\times10^{-2}$, the static torque is always negative and the planet would not become
trapped.

Overall, Fig.~\ref{fig:statictq_iso} is in a very good agreement with Fig.~\ref{fig:at}. The trapping is predicted
for the same values of $\alpha_{\mathrm{MRI}}$ and the trap location differs only by $\simeq$$0.01\,\mathrm{au}$ on average. Although we advocated in Sect.~\ref{sec:migra_vs_alpha} that the planet needs to be migrating
for the Type II torque to be evaluated correctly, we demonstrated here that static torque measurements
are applicable when searching for the existence of the migration trap and its location.

Finally, we check that the surface density distribution and streamlines
behave in a similar manner compared to experiments with a migrating planet,
mainly in the terms of the front-rear asymmetry of the {co-rotation} region.
Fig.~\ref{fig:statictq_dens} depicts the final state of our simulation
with $\alpha_{\mathrm{DZ}}=10^{-3}$, $\alpha_{\mathrm{MRI}}=10^{-1}$
and TOI-216b fixed at $a_{\mathrm{p}}=1.1\,\mathrm{au}$, close to the expected
trap location. By comparing to panel (f) of Fig.~\ref{fig:fiduc_dens_evol},
we immediately see that the disk response to planet-driven perturbations
is qualitatively the same and the torque felt by the planet is inevitably
affected by the overdensity in the front horseshoe region.

\section{Non-isothermal simulations at 0.1 au}
\label{sec:simu_noniso}

The aim of this section is to improve the realism of our simulations
and to support or refine our findings from Sect.~\ref{sec:simu_iso}.
We thus study a non-isothermal disk with its thermodynamics driven by compressional,
viscous and irradiative heating balanced by simplified vertical radiative cooling (Sect.~\ref{sec:model_noniso}).
At the same time, we shift the inner disk rim inwards, to a location close to the 
current orbit of TOI-216b, $r_{\mathrm{MRI}}=0.11\,\mathrm{au}$ \citep{Nesvorny_etal_2022ApJ...925...38N}.
The remaining fixed parameters
of our non-isothermal simulations are summarized in Table~\ref{tab:params_noniso}.

\begin{figure}
    \centering
    \includegraphics[width=\columnwidth]{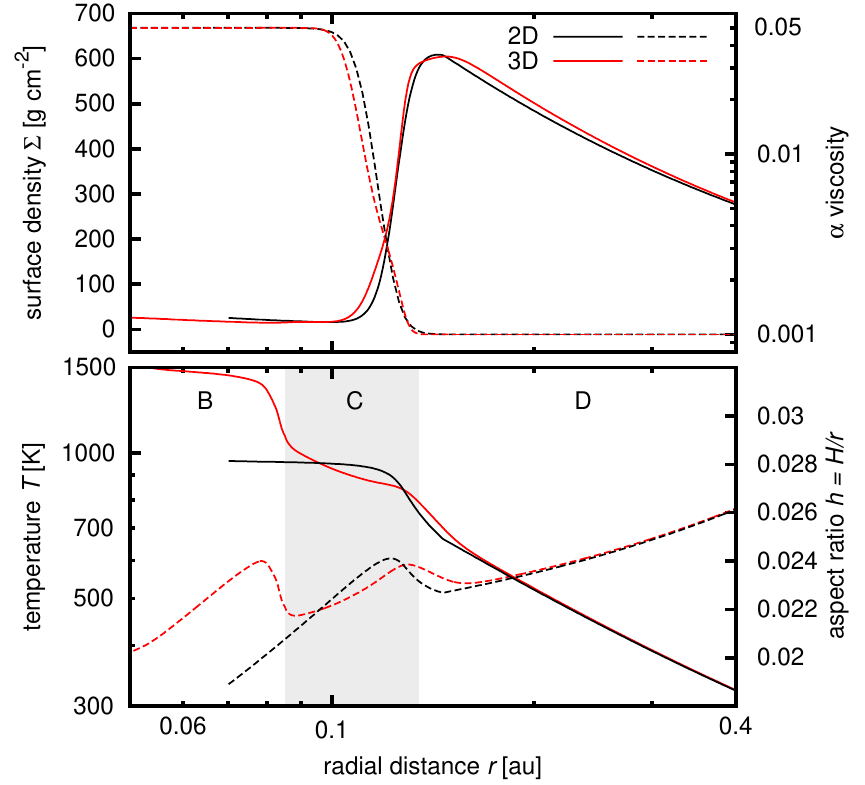}
    \caption{
    Radial profiles of the surface density $\Sigma$ (top, left vertical axis, solid curves),
    $\alpha$ viscosity (top, right vertical axis, dashed curves), temperature $T$ (bottom, left vertical axis, solid curves)
    and aspect ratio $h$ (bottom, right vertical axis, dashed curves). A comparison is shown between our 2D non-isothermal
    model (black curves) and a 3D vertically resolved radiative hydrostatic model (red curves).
    The bottom panel is split into three regions discussed in Sect.~\ref{sec:model_noniso}, namely
    the dust halo (B), the condensation front (C) and the flared disk (D) \citep[e.g.][]{Ueda_etal_2017ApJ...843...49U}.
}
    \label{fig:noniso_disk}
\end{figure}

\subsection{Model verification}

In order to demonstrate the reliability of our 2D non-isothermal model,
we compared the equilibrium radial profiles of several characteristic quantities
to those found in vertically resolved 3D simulations with radiation transfer.
The comparison is shown in Fig.~\ref{fig:noniso_disk}. The displayed
case is for $\dot{M}=10^{-9}\,M_{\odot}\,\mathrm{yr}^{-1}$,
$\alpha_{\mathrm{DZ}}=10^{-3}$, $\alpha_{\mathrm{MRI}}=5\times10^{-2}$.
The reference 3D simulation in Fig.~\ref{fig:noniso_disk}
is based on the radiative hydrostatic approach of \cite{Flock_etal_2016ApJ...827..144F,Flock_etal_2019A&A...630A.147F},
which we have recently re-implemented in Fargo3D (Chrenko et al. in prep.).
The viscosity transition in the 3D reference simulation is temperature-dependent,
following
\begin{equation}
\alpha(T)=\frac{1}{2}(\alpha_{\mathrm{MRI}}-\alpha_{\mathrm{DZ}})\left(1-\tanh{\frac{900\,\mathrm{K}-T}{25\,\mathrm{K}}}\right) + \alpha_{\mathrm{DZ}} \, .
\label{eq:alpha_Tdep}
\end{equation}

To model the radius-dependent $\alpha$-transition in the 2D simulation (see Eq.~\ref{eq:transition}), we chose $r_{\mathrm{MRI}}=0.11\,\mathrm{au}$ together with $\Delta r_{\mathrm{MRI}}=2.7H\simeq0.007\,\mathrm{au}$ and Fig.~\ref{fig:noniso_disk} (top panel) shows that the resulting profile approximates the $\alpha(T)$ profile of the reference 3D simulation well enough.
Consequently, the profile of $\Sigma(r)$ is also recovered with good accuracy. Let us also point out that 
the fiducial width of the viscosity transition used in Sect.~\ref{sec:simu_iso}, $\Delta r_{\mathrm{MRI}}=2H$,
was slightly narrower than we assume here but still relatively similar.

As for the temperature profile (which directly sets the aspect ratio profile),
bottom panel of Fig.~\ref{fig:noniso_disk} reveals that we obtain
the correct radial dependence
in the optically thick flared disk (region D) as well as the steep increase and the plateau 
near the rounded off evaporation front of silicates (region C).
The inner part of the temperature plateau is too extended but this is because we do not attempt to model 
the dust halo (region B in Fig.~\ref{fig:noniso_disk}) with our 2D non-isothermal model (as explained in Sect.~\ref{sec:model_noniso}).
This inconsistency does not influence our results because whether the planet becomes trapped
at the viscosity transition or not is decided at larger radii.

\subsection{Parametric study of the static torque}

\begin{figure*}
    \centering
    \begin{tabular}{ccc}
        \includegraphics[width=0.31\textwidth]{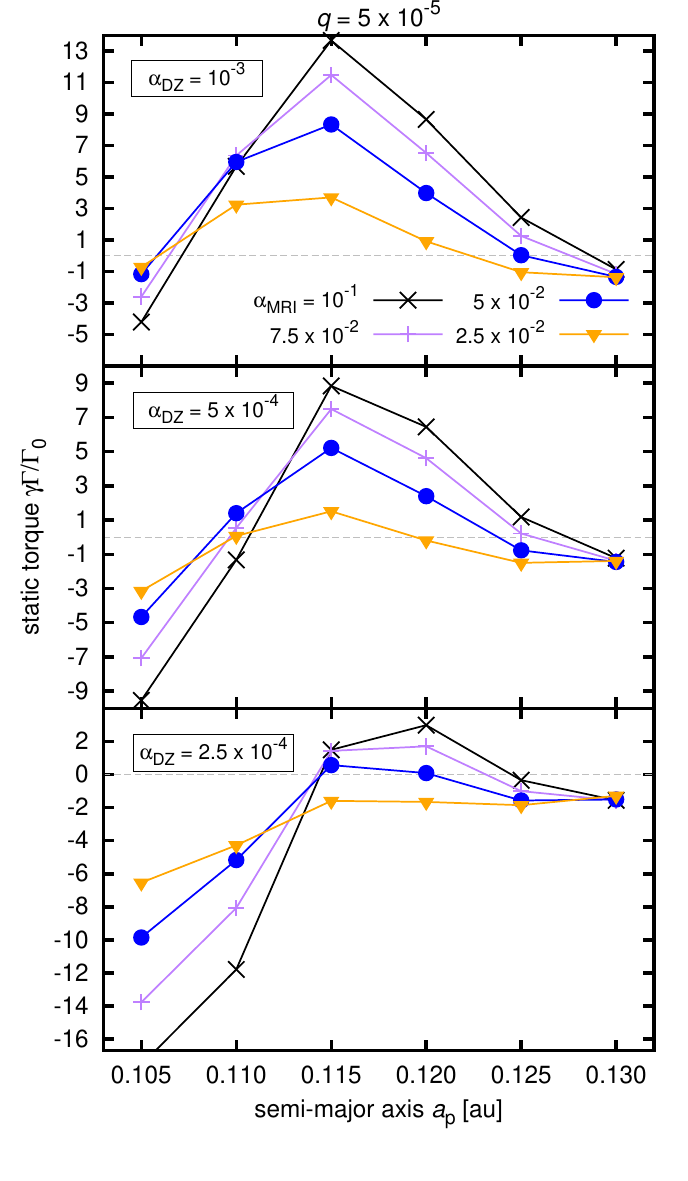} &
        \includegraphics[width=0.31\textwidth]{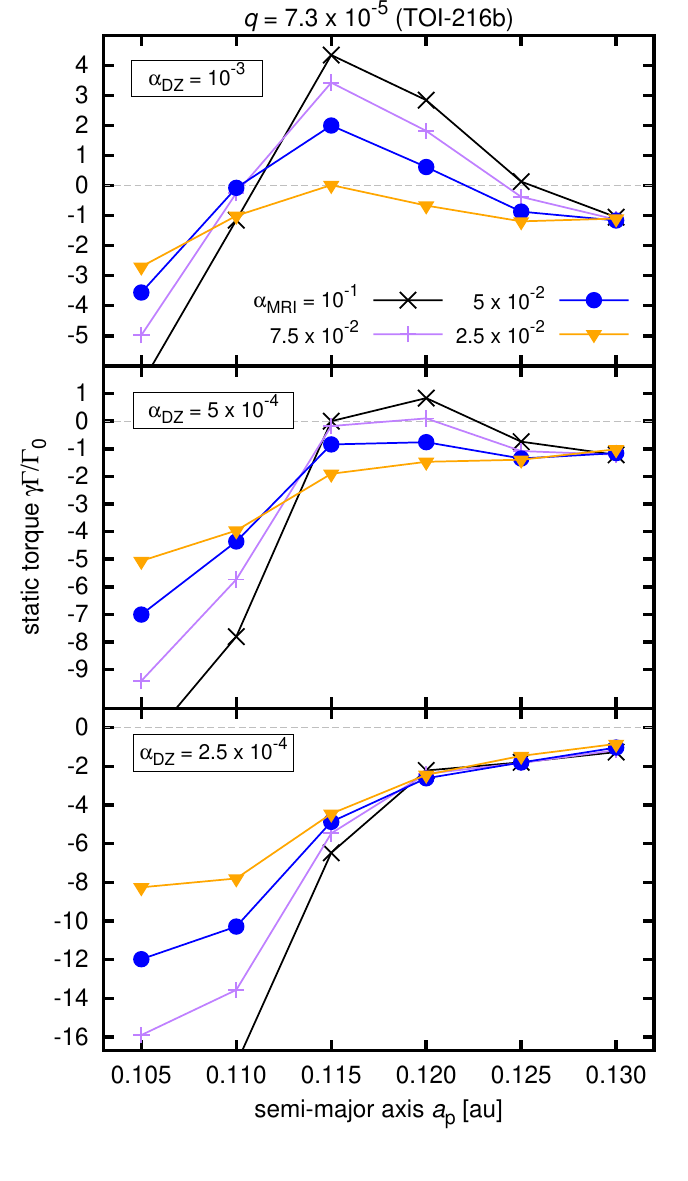} & \includegraphics[width=0.31\textwidth]{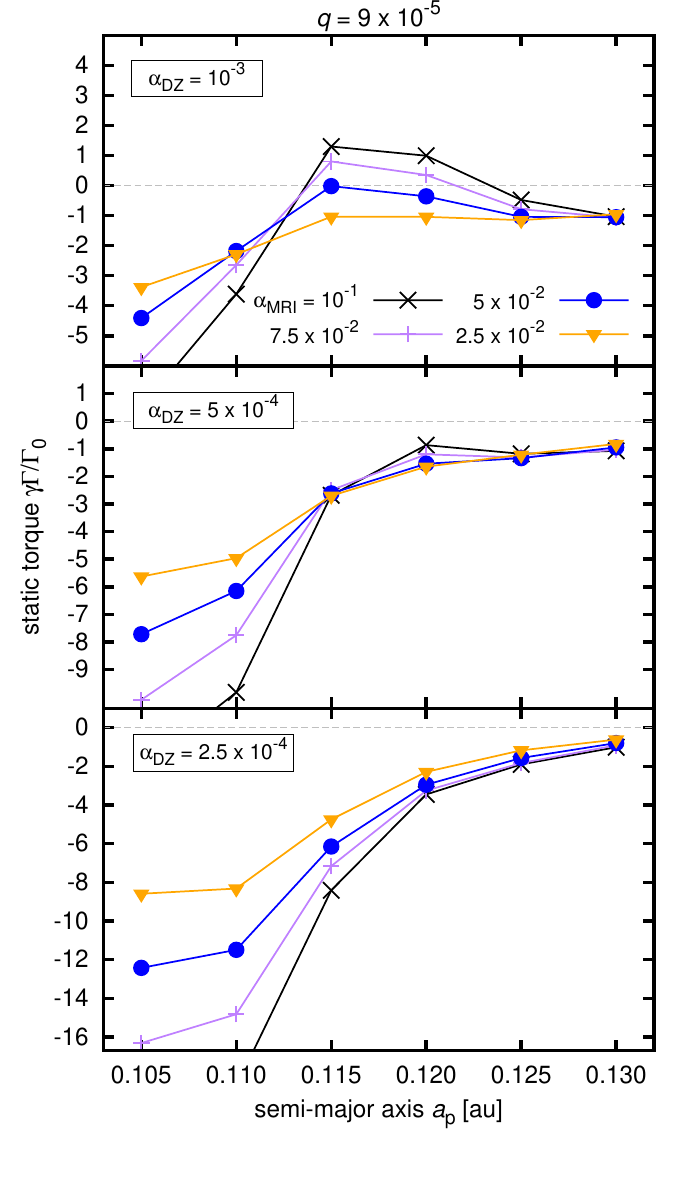} \\
    \end{tabular}
    \caption{Normalized static torque $\gamma\Gamma/\Gamma_{0}$ 
    measured in our non-isothermal simulations as a function of the planetary 
    orbital distance $a_{\mathrm{p}}$. We show measurements for $q=5\times10^{-5}$ (left column), $7.3\times10^{-5}$ (TOI-216b; middle column)
    and $9\times10^{-5}$ (right column). Individual rows correspond to different viscosities $\alpha_{\mathrm{DZ}}$
    in the dead zone (see the boxed labels) and colour-coded points with line segments correspond 
    to different viscosities $\alpha_{\mathrm{MRI}}$ in the MRI-active zone (see the plot legend).
    The horizontal dashed line is where the total torque is zero.
    We point out that the extent of the vertical axis in the left column differs from the middle
    and right columns.
    }
    \label{fig:noniso_statictq}
\end{figure*}

\begin{figure}
    \centering
        \includegraphics[width=\columnwidth]{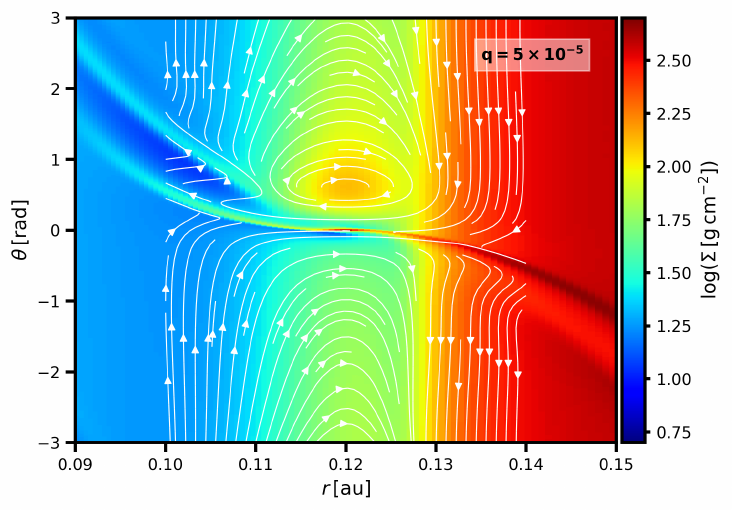}
        \includegraphics[width=\columnwidth]{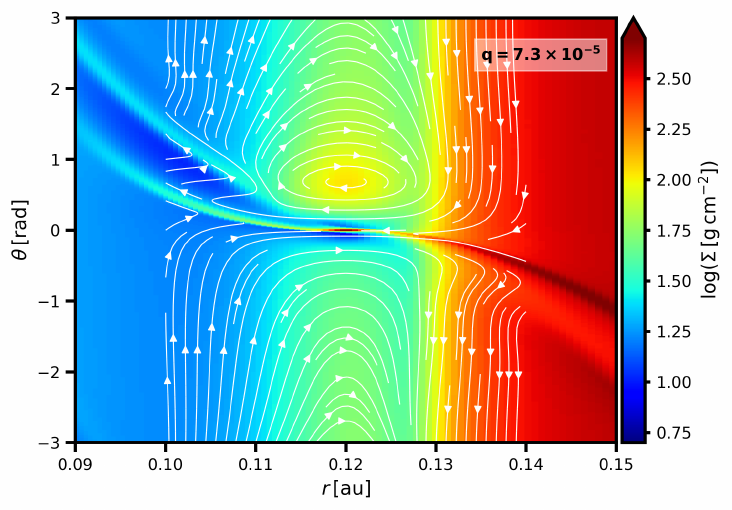}
        \includegraphics[width=\columnwidth]{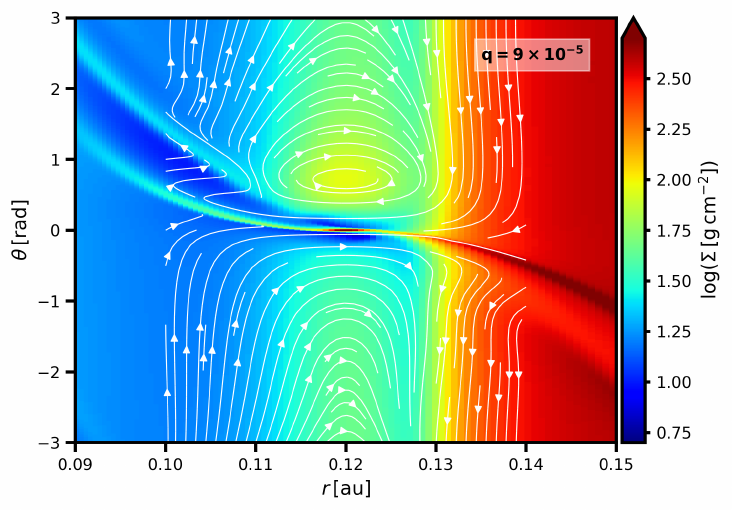}
    \caption{Logarithm of the gas surface density in non-isothermal
    simulations with $\alpha_{\mathrm{DZ}}=10^{-3}$ and $\alpha_{\mathrm{MRI}}=5\times10^{-2}$. The panels correspond
    to different planet-to-star mass ratios as marked by the labels.
    Each planet is held in a fixed circular orbit with $a_{\mathrm{p}}=0.12\,\mathrm{au}$. We overplot streamlines of the gas flow
    relative to the planet with white curves. The figure is also available as an online \texttt{movie} that scans all values of $a_{\mathrm{p}}$ for which the static
    torque was measured.
    Corresponding torque measurements can be found in the top row of Fig.~\ref{fig:noniso_statictq}
    (blue data points).
    }
    \label{fig:noniso_gasdens}
\end{figure}

\begin{figure}
    \centering
        \includegraphics[width=0.8\columnwidth]{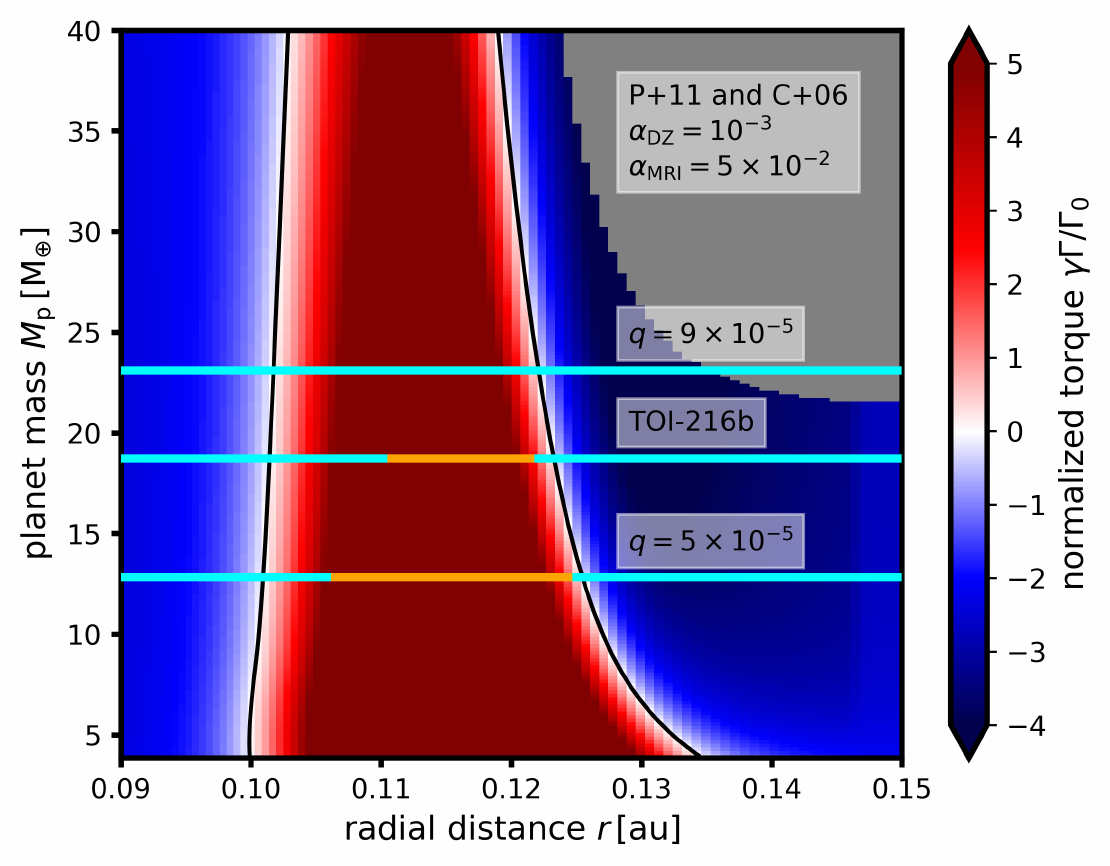}
        \includegraphics[width=0.8\columnwidth]{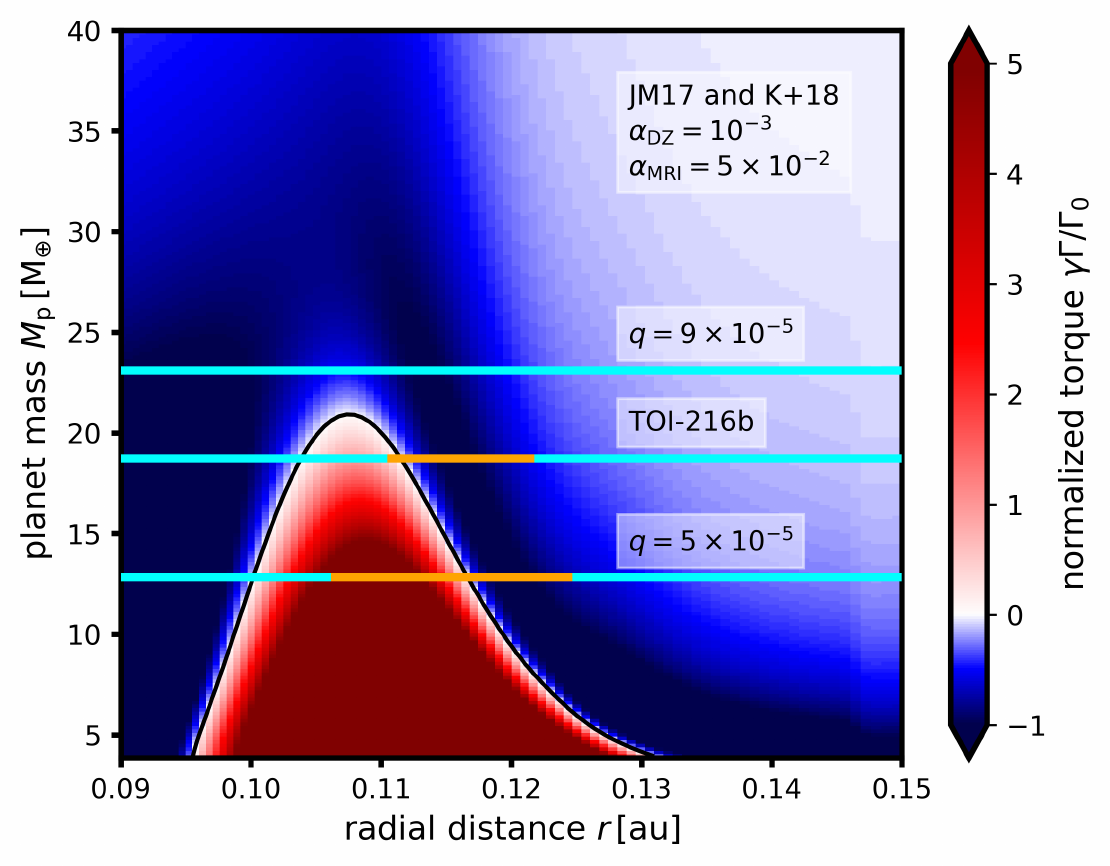} \includegraphics[width=0.8\columnwidth]{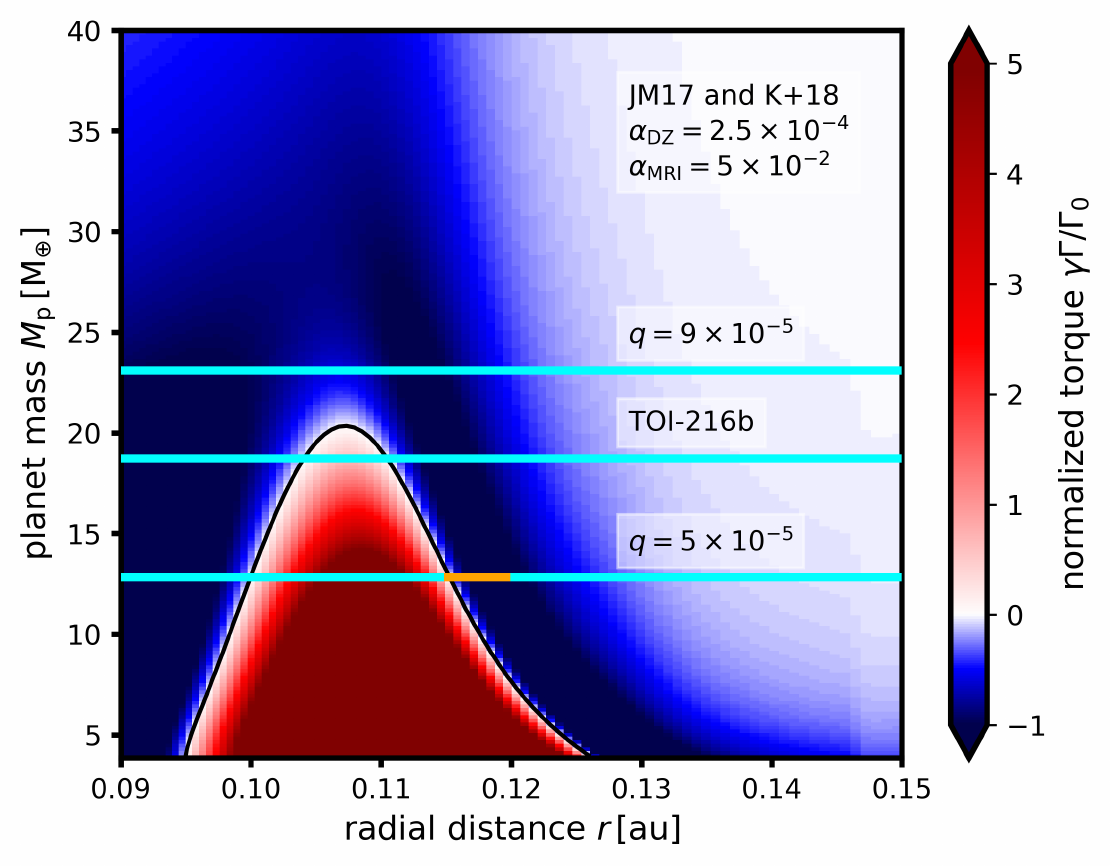}
    \caption{Qualitative comparison between our hydrodynamic simulations (horizontal lines)
    and torque formulae (colour gradient) that are often used in N-body simulations of planet formation.
    Migration maps consisting of red and blue colour gradients mark the regions of the parametric space in which the torque formulae predict outward and inward migration, respectively.
    The black isocontour is where the predicted torque is exactly zero.
    Orange and cyan lines mark the extent of semi-major axes for which our simulations resulted in positive and negative torques, respectively (see Fig.~\ref{fig:noniso_statictq}).
    In the top panel, $\gamma\Gamma/\Gamma_{0}$ is calculated following
    \cite{Paardekooper_etal_2011MNRAS.410..293P} for all planetary masses that do not violate the gap-opening criterion of \cite{Crida_etal_2006Icar..181..587C} (planets that do violate Crida's criterion belong to the grey region). In middle and bottom panels, the torque is calculated
    according to \cite{Jimenez_Masset_2017MNRAS.471.4917J} and blended between Type I and Type II regimes
    using the recipe of \cite{Kanagawa_etal_2018ApJ...861..140K}.
    Top and middle panels are based on the unperturbed radial profiles of our non-isothermal
    disk with the viscosity transition $\alpha_{\mathrm{DZ}}=10^{-3}$ and $\alpha_{\mathrm{MRI}}=5\times10^{-2}$ (see Fig.~\ref{fig:noniso_disk});
    viscosities in the bottom panel are $\alpha_{\mathrm{DZ}}=2.5\times10^{-4}$ and $\alpha_{\mathrm{MRI}}=5\times10^{-2}$. We recall that the planet mass $M_{\mathrm{p}}$ is obtained from $q$
    assuming the stellar mass of TOI-216, $M_{\star}=0.77\,M_{\odot}$.}
    \label{fig:torque_maps}
\end{figure}

In our non-isothermal simulations, we focus mainly on
static torque measurements motivated by the success of this method in Sect.~\ref{sec:static}.
Due to computational limitations, we opt for a slightly coarser grid resolution compared to Sect.~\ref{sec:simu_iso}---
for TOI-216b ($q=7.3\times10^{-5}$), we resolve the Hill radius and the half-width
of the horseshoe region by 5 and 13 cells, respectively. To explore the free parameters, we combine three 
values of $\alpha_{\mathrm{DZ}}=(2.5,5,10)\times10^{-4}$, four values of $\alpha_{\mathrm{MRI}}=(2.5,5,7.5,10)\times10^{-2}$
and three values of $q=(5,7.3,9)\times10^{-5}$. The orbital radius of the planet varies as $a_{\mathrm{p}}=(0.105,0.11,0.115,0.12,0.125,0.13)\,\mathrm{au}$.
We again scale $\dot{M}$ in the same way as $\alpha_{\mathrm{DZ}}$ but, compared to locally
isothermal simulations, we cannot guarantee an identical surface density profile in the dead zone.
The reason is that modifying $\dot{M}$ leads to a different $Q_{\mathrm{visc}}$ and thus the resulting local temperature becomes slightly different, as well as $\nu$ and $\Sigma$.
However, the differences between individual simulations are rather marginal because $Q_{\mathrm{irr}}$ 
dominates over $Q_{\mathrm{visc}}$ for the chosen parameters.
We introduce the planet over $500\,P_{\mathrm{orb}}(r_{\mathrm{MRI}})$ and the total timescale of each simulation
is $2,500\,P_{\mathrm{orb}}(r_{\mathrm{MRI}})$.

Fig.~\ref{fig:noniso_statictq} summarizes our sweep of the parametric space in terms of the normalized
torque $\gamma\Gamma/\Gamma_{0}$. The migration trap is expected to occur wherever the torque switches
from negative to positive when moving from larger orbital radii inwards. The general trend in 
Fig.~\ref{fig:noniso_statictq}  is such that the trapping becomes less efficient
with increasing $q$ (left to right in Fig.~\ref{fig:noniso_statictq}), decreasing $\alpha_{\mathrm{DZ}}$
(top to bottom in Fig.~\ref{fig:noniso_statictq}) and decreasing $\alpha_{\mathrm{MRI}}$
(when the torque is positive, it has a larger magnitude e.g. for crosses than for triangular data points).
In other words, for parameters that allow opening of a more prominent gap in the dead zone
(i.e. decreasing $\alpha_{\mathrm{DZ}}$ or increasing $q$), a stronger viscosity contrast over the transition zone is required for the trap to operate. If the gap is too deep, the trap
fails even for the largest tested value of $\alpha_{\mathrm{MRI}}=10^{-1}$.

Looking at the case of $q=5\times10^{-5}$, which for $M_{\star}=0.77\,M_{\odot}$ translates to 
$M_{\mathrm{p}}=13\,M_{\oplus}$, we see that it can be trapped relatively easily. The only combination of viscosities that fails to trigger the trap is $\alpha_{\mathrm{DZ}}=2.5\times10^{-4}$ and $\alpha_{\mathrm{MRI}}=2.5\times10^{-2}$. However, based on the general trend described in the previous
paragraph, we can predict that it would get progressively difficult to trap the planet if the dead zone viscosity was lower than considered in our parametric sweep, e.g. $\alpha_{\mathrm{DZ}}\lesssim10^{-4}$.

Focussing now on TOI-216b with $q=7.3\times10^{-5}$, we see that $\alpha_{\mathrm{DZ}}=10^{-3}$ requires
$\alpha_{\mathrm{MRI}}\gtrsim5\times10^{-2}$ in the MRI-active zone to sustain the trap
and $\alpha_{\mathrm{DZ}}=5\times10^{-4}$ leads to trapping only if $\alpha_{\mathrm{MRI}}\gtrsim7.5\times10^{-2}$.
For $\alpha_{\mathrm{DZ}}=2.5\times10^{-4}$, the trap does not exist and the planet would continue
migrating inwards. Since trapping of TOI-216b at the viscosity transition is needed
to explain the system's architecture \citep{Nesvorny_etal_2022ApJ...925...38N},
the results obtained here put constraints
on the viscosity values in the central parts of the natal disk of TOI-216.

For the most massive planet shown in Fig.~\ref{fig:noniso_statictq} with $q=9\times10^{-5}$, only $\alpha_{\mathrm{DZ}}=10^{-3}$ and $\alpha_{\mathrm{MRI}}\gtrsim7.5\times10^{-2}$ 
trap the planet at the inner rim. For $\alpha_{\mathrm{DZ}}\leq5\times10^{-4}$, the trapping
is not possible within our parametric range.

Finally, we investigate the surface density distribution and streamlines.
For a comparison with non-isothermal simulations, we select the viscosity combination
of $\alpha_{\mathrm{DZ}}=10^{-3}$ and $\alpha_{\mathrm{MRI}}=5\times10^{-2}$.
Fig.~\ref{fig:noniso_gasdens} shows the density maps for planets positioned 
at $a_{\mathrm{p}}=0.12\,\mathrm{au}$. We chose this combination of parameters
to (i) provide a straightforward comparison with Fig.~\ref{fig:gasdens_q} and (ii) cover various total torques (from Fig.~\ref{fig:noniso_statictq}, we see
that the torque is large and positive for $q=5\times10^{-5}$,
small and positive for TOI-216b, small and negative for $q=9\times10^{-5}$).

In general, Fig.~\ref{fig:noniso_gasdens} is in accordance with Sect.~\ref{sec:simu_iso}. We recover (i) the front island of librating streamlines,
(ii) the mass accumulation within this island that can counteract the negative Lindblad torque, and (iii) the overall depletion of the co-rotation region with increasing $q$. The only difference that we notice is in the general behaviour of streamlines 
in the co-rotation region---a great part of these streamlines goes from the
inner to the outer disk without encountering the planet.
Nevertheless, we confirm here that the migration trap at the inner disk rim for planets with moderate gaps is caused by the presence of a prominent overdensity located in the front horseshoe region, in accordance with our locally isothermal simulations.

\section{Discussion}
\label{sec:discussion}

\subsection{Implications}

Perhaps the most intriguing implication of our study
is that sub-Neptunes and Neptunes
with $q$ similar to our parametric span
can only be trapped at the inner disk rim for specific combinations
of viscosities in the inner MRI-active zone
and the outer dead zone. Therefore, we obtain hints about processes responsible
for generating turbulent stress in natal disks of exoplanetary systems
that harbour said sub-Neptunes and Neptunes at $\simeq$$0.1\,\mathrm{au}$.
We recall that the trapping is required for both in-situ
\citep[e.g.][]{Chatterjee_Tan_2014ApJ...780...53C,Jankovic_etal_2019MNRAS.484.2296J}
and migratory scenarios of planet formation
\citep[e.g.][]{Ida_Lin_2010ApJ...719..810I,Cossou_etal_2013A&A...553L...2C,Izidoro_etal_2017MNRAS.470.1750I}
because without
the disk torque becoming zero, a concentration of planets
would never be achieved \citep[see also][]{Lambrechts_etal_2019A&A...627A..83L,Venturini_etal_2020A&A...643L...1V,Venturini_etal_2020A&A...644A.174V}.

The constraints that we find are still somewhat broad (for instance, trapping of TOI-261b
requires $\alpha_{\mathrm{MRI}}\gtrsim5\times10^{-2}$ when $\alpha_{\mathrm{DZ}}=10^{-3}$ or $\alpha_{\mathrm{MRI}}\gtrsim7.5\times10^{-2}$ when $\alpha_{\mathrm{DZ}}=5\times10^{-4}$) 
yet they are valuable because (i) they confirm that the MRI activity
in the central part of the disk needs to be relatively large
and (ii) the 'dead zone' cannot become fully dead 
but needs to maintain a considerable residual level of viscosity
(e.g. $\alpha_{\mathrm{DZ}}\geq5\times10^{-4}$ for TOI-216b),
otherwise an extremely large $\alpha_{\mathrm{MRI}}$ would be needed
to trigger the trap and that might be difficult to explain with our current
understanding of viscosity-generating processes.
For comparison purposes, let us recall that \cite{Flock_etal_2017ApJ...835..230F}
find $\alpha_{\mathrm{MRI}}\simeq10^{-1}$ and $\alpha_{\mathrm{DZ}}\simeq10^{-3}$
from MHD models of the inner disk turbulence. These values are consistent
with the existence of the planet trap for TOI-216b and similar planets.

Future works can bring additional constraints in conjunction with our study. If, for
example, future MHD models were to discover that viscosity transitions
are rather wide than narrow, the bottom panel of our Fig.~\ref{fig:at_dr}
would immediately imply $\alpha_{\mathrm{MRI}}\gtrsim10^{-1}$ for TOI-216b.

Finally, since the trapping becomes progressively difficult for $q$
larger than that of TOI-216b, a question arises as to whether there could be a
dynamical contribution to the Neptunian desert.
We speculate that general properties of the viscosity transition at the inner rim 
might be such that sub-Neptunes are trapped easily, Neptunes like TOI-216b 
are trapped occasionally (as their abundance is quite small) and super-Neptunes
are not trapped. Their continuing inward migration would thus contribute to their observational paucity---it would `clear the desert'.
Although the likelihood of this effect cannot be 
assessed at the moment, we merely wish to point it out as a possibility.
If our speculations were true, additional questions would require answers, for instance,
what happens with super-Neptunes once they reach the disk edge carved by the magneto-spheric cavity? Are these planets simply evaporated or could they stop once their outer Lindblad resonances migrate out of the disk? Could they become hot Jupiters by merging with companion super-Neptunes that have migrated in a similar fashion?

\subsection{Relevance for N-body simulations with planet migration}
\label{sec:maps}

Our work has important implications for planet population synthesis
as well as for N-body simulations with prescribed migration.
The latter cannot account for the effects shown in our study
because the influence of the planet on the inner rim structure and the development of the front-rear asymmetry in the co-rotation region have not yet been included in the migration formulae \citep{Paardekooper_etal_2011MNRAS.410..293P,Jimenez_Masset_2017MNRAS.471.4917J,Kanagawa_etal_2018ApJ...861..140K}.
Therefore, N-body simulations that lead to a formation of a sub-Neptune or Neptune
as the innermost planet of the planet chain might suffer from systematic errors
because such an innermost planet would feel an incorrect torque. Our study thus
adds to the argument of \cite{Brasser_etal_2018ApJ...864L...8B} who demonstrate
that N-body models with planet migration in the vicinity of the disk edge
require refinements.

To highlight the incompleteness of torque formulae, we computed
the normalized torque by applying the recipe of \citet[][see their Sect.~5.7]{Paardekooper_etal_2011MNRAS.410..293P}
to the radial profiles of our non-isothermal disk (unperturbed by the planet). We chose
$\alpha_{\mathrm{DZ}}=10^{-3}$ and $\alpha_{\mathrm{MRI}}=5\times10^{-2}$, which is the same disk as in Figs.~\ref{fig:noniso_disk} and \ref{fig:noniso_gasdens}. Since the formulae of \cite{Paardekooper_etal_2011MNRAS.410..293P} are only applicable in the Type I 
regime of planetary migration, a criterion is needed to define gap-opening planets that
migrate in the Type II regime and exclude them from the calculation. One such criterion
is that of \cite{Crida_etal_2006Icar..181..587C}, which 
remains to be used \citep[e.g.][]{Flock_etal_2019A&A...630A.147F,Venturini_etal_2020A&A...644A.174V,Emsenhuber_etal_2021A&A...656A..70E}
even though updated criteria
do exist \citep{Kanagawa_etal_2018ApJ...861..140K,Duffell_2020ApJ...889...16D}.

Combining \cite{Paardekooper_etal_2011MNRAS.410..293P} and \cite{Crida_etal_2006Icar..181..587C}, we obtained the migration map shown in the top panel of Fig.~\ref{fig:torque_maps}. After comparing with the results of our simulations (horizontal
lines in Fig.~\ref{fig:torque_maps}), it is clear that the map
correctly predicts the trap location but it incorrectly
overpredicts the migration trap efficiency---trapping is predicted for planets as massive as $M_{\mathrm{p}}\simeq115\,M_{\oplus}$ or $q=4.5\times10^{-4}$ (more massive than TOI-216b by a factor of six).
This error is caused by the failure of the gap-opening criterion of \cite{Crida_etal_2006Icar..181..587C} to account for shallow gaps and thus the formulae of 
\cite{Paardekooper_etal_2011MNRAS.410..293P} are applied even to planetary masses
for which the disk response is already non-linear.

Next, we tested more recent migration formulae. Specifically, we calculated
the Lindblad torque $\Gamma_{\mathrm{L}}$ and the co-rotation torque $\Gamma_{\mathrm{C}}$
according to \citet[][see their Sect.~5.1]{Jimenez_Masset_2017MNRAS.471.4917J} and then we blended them together
as suggested by \cite{Kanagawa_etal_2018ApJ...861..140K}:
\begin{equation}
    \Gamma = \frac{\Gamma_{\mathrm{L}}+\Gamma_{\mathrm{C}}\exp\left(-K/20\right)}{1+0.04K} \, ,
    \label{eq:kanaga_blend}
\end{equation}
where $K$ is given by Eq.~(\ref{eq:K}). The blending should ensure a smooth transition
between Type I and Type II migration and the resulting torque should be applicable even
when the planet opens a gap.

The result is shown in the middle panel of Fig.~\ref{fig:torque_maps}
and, at first glance, it seems to match our simulations very well---although the 
trap location is shifted inwards by $\simeq$$0.01\,\mathrm{au}$ compared to our simulations,
the maximum
planetary mass that can be trapped is found correctly. However, further experiments 
revealed that the match is rather coincidental. By repeating the calculation for the disk
with $\alpha_{\mathrm{DZ}}=2.5\times10^{-4}$ and $\alpha_{\mathrm{MRI}}=5\times10^{-2}$,
as shown in the bottom panel of Fig.~\ref{fig:torque_maps}, we discover that the 
island of outward migration is much less sensitive to the viscosity contrast compared
to our simulations. While we can barely trap the planet with $q=5\times10^{-5}$ in our simulations with a low-viscosity dead zone, the torque formulae predict a similar trap as in the middle panel of Fig.~\ref{fig:torque_maps} and thus they overestimate the trapping efficiency yet again. Therefore, we conclude that the available torque formulae indeed fail to recover the results found in our hydrodynamic simulations.

\subsection{Comparison to other works}

Our results indicate a different regime of trapping compared to \cite{Masset_etal_2006ApJ...642..478M}, as we already pointed out in previous
sections. In addition to that, let us emphasize that
our mechanism differs from the vortex-aided trapping as well \citep{Regaly_etal_2013MNRAS.433.2626R,Ataiee_etal_2014A&A...572A..61A}
because in that scenario, a Rossby vortex needs to be formed at a viscosity
transition (or at a density bump) prior to the planet reaching the vortex location.
The trap is then triggered during the orbital crossing of the planet and the vortex.
In our case, instead, if a vortex forms, it does so gradually
inside the co-rotation region of the planet. The vortex itself is not
a necessary prerequisite for the trapping, it is rather a secondary result
of gas concentration in the asymmetric co-rotation region.
We also detect vortices forming outside the planet's co-rotation region
but these are not long-lived and do not seem to contribute to trapping.

We also find differences compared to \cite{Romanova_etal_2019MNRAS.485.2666R},
likely because the authors did not account for viscosity transitions, their simulation
timescales were short (did not allow for full gap opening), the whole planet
mass was inserted immediately in the simulation and the planet was allowed
to migrate straightaway. On the other hand, \cite{Romanova_etal_2019MNRAS.485.2666R}
used more realistic 3D simulations and it is left for future work
to verify whether our findings can be recovered in 3D as well.

Unlike in \cite{Faure_Nelson_2016A&A...586A.105F}, our non-isothermal simulations do not form any RWI vortices
at the transition between the dead and active zones that would undergo cyclic
evolution and migration. We think that some of our model features would need
to be modified in order to recover results of \cite{Faure_Nelson_2016A&A...586A.105F},
for example, the viscosity would need to be temperature-dependent (not radially dependent), 
our dead zone would need to have zero viscosity,
or our aspect ratio would need to be larger.

A firm overlap can be found between our study and \cite{Hu_etal_2016ApJ...816...19H}
who explore the migration trap at the inner disk rim in the context
of the inside-out planet formation of close-in exoplanetary systems \citep{Chatterjee_Tan_2014ApJ...780...53C}.
By looking at the radial surface density profile perturbed by their most
massive planet with $q\approx3.5\times10^{-5}$ (see the top panel of their Fig. 3),
we see that it roughly corresponds to the perturbation produced by our least massive planet
with $q=1.5\times10^{-5}$ (see panel (a) of our Fig.~\ref{fig:sigma1d_q}). The
slight difference in the gap-opening ability is likely caused by differences
in the aspect ratio of the background disk model. 
However, one effect of \cite{Hu_etal_2016ApJ...816...19H} that is not considered in our
model is the outward shift of the dead-zone inner edge related to the penetration
of ionizing X-rays into the gaped disk.

Finally, let us discuss the streamline topology. The appearance of a front island
of librating streamlines is reminiscent of results of \cite{Ogilvie_Lubow_2006MNRAS.370..784O} (see their Fig.~4) where the deformation
of the co-rotation streamlines is a result of planet migration. The front-rear
asymmetry is additionally reminiscent of Type III migration \citep{Masset_Papaloizou_2003ApJ...588..494M,Peplinski_etal_2008MNRAS.386..164P},  although that regime is usually considered in the context of runaway inward migration \citep[cf.][]{Peplinski_etal_III_2008MNRAS.387.1063P}
and not in the context of planet traps.
We point out that in our case, compared to studies mentioned above, 
the asymmetry of the co-rotation region is recovered regardless of planet migration, in both static and dynamical numerical experiments.
Therefore, as stated in \cite{Paardekooper_2014MNRAS.444.2031P}, the formation of a one-sided mass deficit in the co-rotation region that we reported
is likely a natural consequence
of the planet finding itself on an initially steep density gradient (the
planet is radially sandwiched between a low-denity/high-viscosity inner region and a high-density/low-viscosity outer region). 
For completeness, let us note that our streamline topology is azimuthally antisymmetric to the one found in \cite{McNally_etal_2017MNRAS.472.1565M} (e.g. their
Figs. 4 and 6).

\subsection{Limitations and future work}

Due to the need to perform
simulations over relatively long timescales, our disk model inevitably contains several simplifications
and some of them should be tested in future to capture the
physics of the inner disk rim in its entire complexity.
The main two drawbacks are (i) the 2D approximation and (ii) the viscosity prescription.
While it is relatively easy to convert our model to 3D (at least the locally
isothermal one), 
improving the viscosity is not straightforward. 
One can at least think of a temperature-dependent viscosity, as already
pointed out, but an equally important question is what happens
with the disk turbulence in the presence of the planetary gap.
If the gap becomes deep in the dead zone (before approaching the inner
rim), can the gap centre become exposed and ionized to increase the turbulence level
locally? And, since our viscous disk is laminar, what would be the evolution in a truly turbulent disk? Only future works can answer these questions.

Regarding the planet itself, we assumed that its mass remains constant. In other words, we neglected
pebble accretion and gas accretion. While the omission of the former can be advocated because the 
pebble isolation mass is typically low in the inner disk \citep[e.g.][]{Bitsch_2019A&A...630A..51B,Venturini_etal_2020A&A...644A.174V}, gas accretion
could still be ongoing. We decided not to include gas accretion in our global 2D models because of two obstacles---the implementation would have to rely on accretion rates adopted from other studies
\citep[e.g.][]{Machida_etal_2010MNRAS.405.1227M,Bodenheimer_Lissauer_2014ApJ...791..103B}
and the planet would have to act like a sink particle \citep[while in an ideal situation, one would like to resolve a 3D gas flow all the way to the planetary envelope; e.g.][]{Lambrechts_Lega_2017A&A...606A.146L,Schulik_etal_2019A&A...632A.118S}.
Nevertheless, gas accretion should be considered in future because it can change the gap profile \citep{Bergez-Casalou_etal_2020A&A...643A.133B}
and thus affect the asymmetry of the co-rotation region as well.

As a side result, we find Type III migration for $q=(1.5$--$4.5)\times10^{-5}$
shortly before these planets reach the viscosity transition.
The importance of this migration type
for the assembly of close-in exoplanetary systems should thus be reassessed
in future studies.

\section{Conclusions}
\label{sec:conclusions}

Motivated by recent claims that the innermost sub-Neptune-like
exoplanets in multi-planetary systems cluster at $0.1\,\mathrm{au}$ \citep{Mulders_etal_2018AJ....156...24M}
and that such clustering is caused by the migration trap acting
at a viscosity (or density) transition at the inner disk rim \citep{Masset_etal_2006ApJ...642..478M,Flock_etal_2019A&A...630A.147F},
we performed 2D locally isothermal and non-isothermal simulations
to revisit the trap mechanism for sub-Neptunes and Neptunes
(although our actual planet mass interval extended beyond these types
of planets as well).
Our working hypothesis was that close-in sub-Neptunes and Neptunes are actually
able to open prominent gaps because of the very low aspect ratio $h\simeq0.02$--$0.03$
that is expected at the disk rim \citep{Flock_etal_2019A&A...630A.147F,Nesvorny_etal_2022ApJ...925...38N}.
Consequently, the migration trap of \cite{Masset_etal_2006ApJ...642..478M} might
fail because it relies on the positive vortensity-driven
co-rotation torque which might be suppressed when the co-rotation region becomes
emptied due to a gap opening.
As our reference
representative of close-in Neptunes, we chose
TOI-216b since its orbital properties and its firm 2:1 resonance with TOI-216c
can be explained in a scenario that relies on the migration trap at the inner rim \citep{Nesvorny_etal_2022ApJ...925...38N}.

We discover a new regime of the migration trap
that is facilitated by the development of a front-rear asymmetry in the co-rotation
(or horseshoe) region of the planet. The front horseshoe region develops
an island of librating streamlines that accumulate a gas overdensity.
Since the gas overdensity is leading the orbital motion of the planet,
it exerts a positive torque and it can balance the negative torque of spiral wakes,
thus halting the migration.
In our locally isothermal runs, we find that the overdensity can turn into a Rossby
vortex if the viscosity contrast is large and if $q\geq6\times10^{-5}$. The trap,
however, persists even in the presence of the vortex.

For the parameters that we explored, the new regime of the migration trap operates
roughly for planet-to-star mass ratios $q\simeq(4$--$8)\times10^{-5}$.
In this mass range, the planets open moderate gaps (for $h\simeq0.023$ that we considered) and they locally erase the initial surface density transition, replacing it
with a bump-like surface density distribution centred on the planet's co-rotation.
For lower $q$, the gaps remain shallow and the trapping proceeds as suggested
by \cite{Masset_etal_2006ApJ...642..478M}. For higher $q$, the total gas
mass in the co-rotation region is further reduced and its net gravitational
torque is no longer capable of sustaining the trap.

Instead of varying the planet mass, similar effects are found if the planet
mass is fixed and the disk viscosity is varied instead. This is because
the gap-opening ability increases with decreasing viscosity.
By studying migration of TOI-216b across viscosity transitions with different
dead-zone viscosities $\alpha_{\mathrm{DZ}}$ and MRI-active viscosities $\alpha_{\mathrm{MRI}}$, we are able to find constraints for their combinations
that promote the trapping. In summary, when $\alpha_{\mathrm{DZ}}=10^{-3}$, TOI-216b can only be trapped for $\alpha_{\mathrm{MRI}}\gtrsim5\times10^{-2}$. Similarly, $\alpha_{\mathrm{DZ}}=5\times10^{-4}$ requires $\alpha_{\mathrm{MRI}}\gtrsim7.5\times10^{-2}$ for the trap to operate.
Finally, the trap never occurs for $\alpha_{\mathrm{DZ}}\leq2.5\times10^{-4}$,
at least for the considered range of $\alpha_{\mathrm{MRI}}$ that had an upper limit of $10^{-1}$. Our results agree between the locally isothermal and
non-isothermal disks, which makes them robust and independent of the disk
thermodynamics, although the viability of some of our model assumptions (see Sect.~\ref{sec:discussion}) should be tested in future studies.
Additionally, we find very similar locations of the migration trap
when comparing dynamical simulations to static simulations with a fixed planetary
orbit. In other words, the existence of the trap does not seem
to strongly depend on the migration history of the planet.

Our findings imply that by studying sub-Neptunes and Neptunes whose
formation requires a migration trap at the inner rim,
it is possible to place constraints
on the viscosity in central parts of their natal protoplanetary disks.
We also speculate that there might be a dynamical contribution to
the existence of the Neptunian desert because
planets more massive than Neptunes are difficult to trap in general.
Finally, we point out that N-body models with prescribed migration
might suffer from systematic errors at the inner rim
because the available torque formulae seem to overpredict
the maximum planet mass that can be trapped (see Sect.~\ref{sec:maps}).

\begin{acknowledgements}
This work was supported by
the Czech Science Foundation (grant 21-23067M)
and the Ministry of Education, Youth and Sports of
the Czech Republic through the e-INFRA CZ (ID:90140, LM2018140).
The work of O.C. was supported by
the Charles University Research program (No. UNCE/SCI/023).
M.F. was  supported by the European Research Council (ERC) project
under the European Union’s Horizon 2020 research and innovation programme
number 757957.
We wish to thank an anonymous referee whose valuable
comments allowed us to improve this paper.
\end{acknowledgements}

%
%

\bibliographystyle{aa}
\bibliography{references}

\begin{appendix}

\section{Viscous heating term}
\label{sec:app_visc}

The viscous dissipation term in full 3D spherical coordinates is \citep[e.g.][]{Mihalas_WeibelMihalas_1984frh..book.....M,DAngelo_Bodenheimer_2013ApJ...778...77D}
\begin{equation}
    Q_{\mathrm{visc,3D}} = \frac{1}{2\nu\rho}\left(\tau_{rr}^{2}+\tau_{\theta\theta}^{2}+\tau_{\phi\phi}^{2}+2\tau_{r\theta}+2\tau_{r\phi}+2\tau_{\theta\phi}\right) \, ,
\end{equation}
where $\tau_{ij}$ are the components of the viscous stress tensor.
Using vertical averaging and omitting zero terms in 2D, we obtain
\begin{equation}
    Q_{\mathrm{visc}} = \frac{1}{2\nu\Sigma}\left(\tau_{rr}^{2}+\tau_{\theta\theta}^{2}+\tau_{\phi\phi}^{2}+2\tau_{r\theta}\right) \, .
\end{equation}
The remaining vertical component is non-zero because in
\begin{equation}
    \tau_{\phi\phi} = 2\nu\Sigma\left(D_{\phi\phi}-\frac{1}{3}\nabla\cdot\vec{v}\right) \, ,
\end{equation}
only the strain tensor component $D_{\phi\phi}$ is zero. Thus the final form for the viscous heating
term is
\begin{equation}
    Q_{\mathrm{visc}} = \frac{1}{2\nu\Sigma}\left(\tau_{rr}^{2}+\tau_{\theta\theta}^{2}+2\tau_{r\theta}\right)+\frac{2\nu\Sigma}{9}\left(\nabla\cdot\vec{v}\right)^{2} \, .
    \label{eq:qvisc}
\end{equation}

\section{Stellar irradiation term}
\label{sec:app_irr}

Irradiating rays propagate from the star and they are impinging
the $\tau\approx1$ surface of the disk with the total flux of
\begin{equation}
    F_{\mathrm{irr}} = \sin{\alpha}\frac{L_{\star}}{4\pi r^{2}} \, ,
    \label{eq:flux_irr}
\end{equation}
where $\alpha$ is the grazing angle and $L_{\star}$ is the stellar luminosity.
Eq.~(\ref{eq:flux_irr}) is applicable when the whole star is visible from an arbitrary point
of the irradiated surface. This is consistent 
with the assumption of an optically thin region inwards from the condensation
front of dust grains\footnote{For comparison, we note that the classical model
of \cite{Chiang_Goldreich_1997ApJ...490..368C} assumes a disk that stretches all 
the way to the central star and thus only the upper stellar hemisphere is visible
from an arbitrary point on the disk surface.}
\citep{Dullemond_etal_2001ApJ...560..957D}.

We assume that the $\tau\approx1$ surface is caused by dust grains
and we write for the energy balance of a dust grain
\begin{equation}
    \frac{L_{\star}}{4\pi r^{2}}\left[\frac{4R_{\star}}{3\pi r} + \frac{H_{\mathrm{ph}}}{r}\left(\frac{\mathrm{d}\ln{H_{\mathrm{ph}}}}{\mathrm{d}\ln{r}}-1\right)\right] = \left(\frac{C_{\mathrm{bw}}}{4}\right)^{-1}\epsilon\sigma_{\mathrm{B}} T_{\mathrm{s}}^{4} \, ,
    \label{eq:grain-flux_balance}
\end{equation}
where $H_{\mathrm{ph}}$ is the photospheric height above the midplane, $C_{\mathrm{bw}}=4$
is the backwarming factor \citep{Kama_etal_2009A&A...506.1199K}, 
and we simplified the grazing angle prescription \citep[cf.][]{Chiang_etal_2001ApJ...547.1077C}.
Close to the star, the first geometrical term in the square brackets dominates and,
using $L_{\star}=4\pi R_{\star}^{2}\sigma_{\mathrm{B}}T_{\star}^{4}$, we obtain the surface temperature
\begin{equation}
    T_{\mathrm{s,in}} = \left(\frac{4}{3\pi\epsilon}\right)^{1/4}\left(\frac{R_{\star}}{r}\right)^{3/4}T_{\star} \, .
    \label{eq:T_surf_in}
\end{equation}
Farther out, the second geometrical term in the square brackets prevails.
To find a solution for the surface temperature, we assume that the disk 
obeys the flaring disk principle with the equilibrium slope
$\mathrm{d}\ln{H_{\mathrm{ph}}}/\mathrm{d}\ln{r}=9/7$ \citep{Chiang_Goldreich_1997ApJ...490..368C}.
The photospheric height is by a factor of few larger than the pressure scale height,
$H_{\mathrm{ph}}=f_{\mathrm{ph}}H$, and this leads to
\begin{equation}
    T_{\mathrm{s,out}}^{4} = \frac{1}{\epsilon}\left(\frac{R_{\star}}{r}\right)^{2}\frac{2f_{\mathrm{ph}}}{7r}HT_{\star}^{4} \, .
    \label{eq:T_surf_out1}
\end{equation}
To remove the dependency on $H$, we take
\begin{equation}
    H = \frac{c_{\mathrm{s}}}{\sqrt{\gamma}\Omega}=\sqrt{\frac{\gamma P/\rho}{\gamma GM_{\star}/r^{3}}}=\sqrt{\frac{\left(\gamma-1\right)c_{V}T_{\mathrm{i,out}}r^{3}}{GM_{\star}}} \, ,
    \label{eq:H_irr}
\end{equation}
with the interior temperature $T_{\mathrm{i,out}}=2^{-1/4}T_{\mathrm{s,out}}$. The latter relation between
the interior and surface temperature is based on the assumption of perfect splitting of
the irradiating flux in halves: one half is radiated away and one half goes deeper into the disk \citep{Dullemond_etal_2001ApJ...560..957D}. Finally,
\begin{equation}
    T_{\mathrm{s,out}} \simeq \left(0.262\frac{f_{\mathrm{ph}}}{\epsilon}R_{\star}^{2}\sqrt{\frac{\left(\gamma-1\right)c_{V}}{GM_{\star}r^{3}}}T_{\star}^{4}\right)^{2/7} \, .
    \label{eq:T_surf_out2}
\end{equation}
A generalized solution for 
the midplane temperature of an optically thick irradiated disk can be written as \citep[see also][]{Ueda_etal_2017ApJ...843...49U}
\begin{equation}
    T_{\mathrm{thick}} = \left[\frac{1}{2}\left(T_{\mathrm{s,in}}^{4} + T_{\mathrm{s,out}}^{4}\right)\right]^{1/4} \, ,
    \label{eq:Tthick_appendix}
\end{equation}
where the factor 1/2 again accounts for the flux splitting.

\section{Supplementary figures}
\label{sec:app_add}

\begin{figure}
    \centering
        \includegraphics[width=\columnwidth]{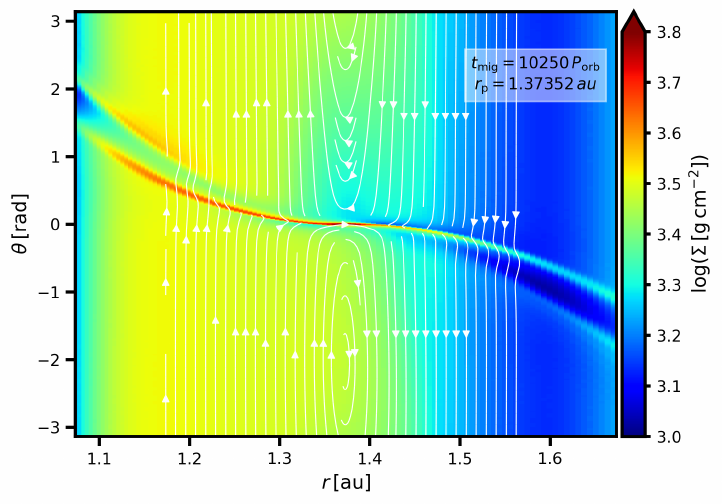}
    \caption{Surface density and streamlines for a migrating 
    super-Earth-sized planet with $q=1.5\times10^{-5}$ at $t_{\mathrm{mig}}=300\,P_{\mathrm{orb}}$
    after the planet was released. The corresponding migration track is shown in bottom panel of Fig.~\ref{fig:at_q} ($\alpha_{\mathrm{DZ}}=10^{-3}$ and $\alpha_{\mathrm{MRI}}=2.5\times10^{-2}$).
    There is an excess mass in the rear horseshoe region
    and the planet thus migrates in the runaway Type III regime.
    }
    \label{fig:peculiar1}
\end{figure}

In this section, we present additional figures to support our claims related
to Sect.~\ref{sec:vary_mass}, Fig.~\ref{fig:at_q}
and the peculiar cases of migration reported therein.
First, we pointed out that planets with $q\leq4.5\times10^{-5}$ undergo
runaway inward Type III migration before becoming trapped
at the viscosity transition. Fig.~\ref{fig:peculiar1}
shows the density distribution and streamlines near the planet
with $q=1.5\times10^{-5}$ migrating through the dead zone with
$\alpha_{\mathrm{DZ}}=10^{-3}$ and approaching the MRI-active
zone with $\alpha_{\mathrm{MRI}}=2.5\times10^{-2}$. 
The displayed state corresponds to $t_{\mathrm{mig}}=300\,P_{\mathrm{orb}}$
in bottom panel of Fig.~\ref{fig:at_q} and we selected the case
in which the Hill cut of the disk material for the torque evaluation was considered.
We can see that there is a clear gas mass deficit in the front horseshoe
region and thus the negative torque felt by the planet is enhanced.
This explains why the planet migrates rapidly towards the disk edge.

\begin{figure}
    \centering
        \includegraphics[width=\columnwidth]{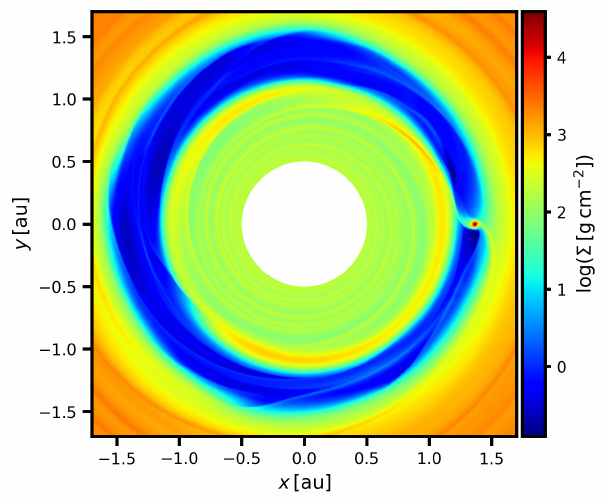}
    \caption{Surface density for a migrating 
    planet with $q=3.6\times10^{-4}$ shortly before the end of the simulation.
    The corresponding migration track is shown in the bottom panel of Fig.~\ref{fig:at_q}.
    The planet is strongly eccentric ($e\simeq0.05$), which manifests itself via the eccentricity of the gap
    and the relative position of the planet with respect to gap edges.
    We point out that this unusual result (compared to our remaining simulations)
    might be caused by a numerical artefact.
    }
    \label{fig:peculiar2}
\end{figure}

Fig.~\ref{fig:peculiar2} shows the case with $q=3.6\times10^{-4}$ (the other
parameters remain the same) for which the planetary migration stagnated 
even before reaching the viscosity transition. The stagnation occurs
due to the excited eccentricity of the planet, $e\simeq0.05\sim2h$, because
the planet starts to encounter the gap walls \citep{Duffell_Chiang_2015ApJ...812...94D}.
However, this case should be approached with caution as it might suffer from numerical
artefacts. The reason is that for the given combination of $q$, $h$ and $\alpha_{\mathrm{DZ}}$, the planet gets very efficient in perturbing the disk.
Consequently, the outer part of the disk becomes eccentric first.
The eccentric disk then starts to wobble
around the domain origin (because the system barycentre becomes offset from the star)
and, because we do account for the indirect terms of the gravitational potential,
the wobble is sustained.
But at the same time, the wobbling disk becomes inconsistent with our boundary conditions that are supplemented with wave-killing zones. These wave-killing zones
essentially enforce axial symmetry close to the domain boundary that does
not match the eccentric disk. We think that future work is needed
in order to exclude the possibility of a nonphysical feedback loop 
between the eccentric disk, a gap-opening
planet with a deep gap and our boundary conditions.

\end{appendix}

\end{document}